\documentclass{aa}  

\usepackage{subfigure}
\usepackage{multirow}
\usepackage{courier}
\usepackage{graphicx}
\usepackage{grffile} 
\usepackage{txfonts}
\usepackage{natbib}
\usepackage{booktabs,caption,fixltx2e}
\usepackage[flushleft]{threeparttable}
\usepackage{setspace}

\usepackage{amstext}

\begin{document}

   \title{Radio spectral properties and jet duty cycle in the \\ restarted radio galaxy 3C388\thanks{Fits files of the radio maps are available at the CDS via anonymous ftp to cdsarc.u-strasbg.fr (130.79.128.5)
or via http://cdsweb.u-strasbg.fr/cgi-bin/qcat?J/A+A/} }

   \author{M. Brienza\inst{1,2}\fnmsep\thanks{m.brienza@ira.inaf.it}
          \and 
          R. Morganti\inst{3,4}          
          \and
          J. Harwood\inst{7}
          \and
           T. Duchet\inst{5}
          \and
          K. Rajpurohit\inst{1}
          \and
          A. Shulevski\inst{6}
          \and
          M. J. Hardcastle\inst{7}
          \and
          V. Mahatma\inst{7}
          \and
          L. E. H. Godfrey
          \and
          I. Prandoni\inst{2}
          \and 
          T. W. Shimwell\inst{3,9}
          \and 
          H. Intema\inst{8,9}
              }

   \institute{Dipartimento di Fisica e Astronomia, Università di Bologna, via P. Gobetti 93/2, 40129, Bologna, Italy
   \and
   INAF – Istituto di Radioastronomia, Via P. Gobetti 101, 40129, Bologna, Italy
   \and
        ASTRON, the Netherlands Institute for Radio Astronomy, Postbus 2, 7990 AA, Dwingeloo, The Netherlands
   \and
        Kapteyn Astronomical Institute, Rijksuniversiteit Groningen, Landleven 12, 9747 AD Groningen, The Netherlands
        \and 
        University of Orléans, 3A avenue de la Recherche Scientifique, 45071 Orléans cedex 2, France        
        \and
        Anton Pannekoek Institute for Astronomy, University of Amsterdam, Postbus 94249, 1090 GE Amsterdam, The Netherlands       
        \and            
        Centre for Astrophysics Research, School of Physics, Astronomy \& Mathematics, University of Hertfordshire, College Lane, Hatfield, Hertfordshire, AL10 9AB, UK
        \and
        International Centre for Radio Astronomy Research – Curtin University, GPO Box U1987, Perth, WA 6845, Australia
        \and
        Leiden Observatory, Leiden University, Niels Bohrweg 2, NL-2333CA Leiden, the Netherlands
}

\abstract
{Restarted radio galaxies represent a unique tool for investigating the duty cycle of the jet activity in active galactic nuclei (AGN). The radio galaxy 3C388 has long been claimed to be a peculiar example of an AGN with multi-epoch activity because it shows a very sharp discontinuity in the GHz spectral index distribution of its lobes.
}
{We present here for the first time a spatially resolved study of the radio spectrum of 3C388 down to MHz frequencies aimed at investigating the radiative age of the source and constraining its duty cycle.}
{We used new low-frequency observations at 144 MHz performed with the Low Frequency Array and at 350 MHz performed with the Very Large Array that we combined with archival data at higher frequencies (614, 1400, and 4850 MHz).}
{We find that the spectral indices in the lower frequency range, 144-614 MHz, have flatter values ($\rm \alpha_{low}\sim$0.55-1.14) than those observed in the higher frequency range, 1400-4850 MHz, ($\rm \alpha_{high}\sim$0.75-1.57), but they follow the same distribution across the lobes, with a systematic steepening towards the edges. However, the spectral shape throughout the source is not uniform and often deviates from standard models. This suggests that mixing of different particle populations occurs, although it remains difficult to understand whether this is caused by observational limitations (insufficient spatial resolution and/or projection effects) or by the intrinsic presence of multiple particle populations, which might be related to the two different outbursts.}
{Using single-injection radiative models, we compute that the total source age is $\lesssim$80 Myr and that the duty cycle is about $\rm t_{on}/t_{tot}\sim$ 60\%, which is enough to prevent the intracluster medium from cooling, according to X-ray estimates. While to date the radio spectral distribution of 3C388 remains a rare case among radio galaxies, multi-frequency surveys performed with new-generation instruments will 
soon allow us to investigate whether more sources with the same characteristics exist.}

   \keywords{galaxies : active - radio continuum : galaxy - individual: 3C388}

   \maketitle


\section{Introduction}
\label{intro}

Restarted radio galaxies represent one of the clearest indications that radio jets driven by active galactic nuclei (AGN) can be episodic. In these sources we can simultaneously observe remnant lobes produced by a previous phase of jet activity, and a pair of active newly-born jets (see \citealp{saikia2009} for a review). 

These sources offer us a unique opportunity to constrain the jet duty cycle in AGN, that is, the timescales of the jet activity and quiescence \citep{morganti2017}. The simultaneous modelling of the radio spectrum of the old and young lobes provides, indeed, an estimate of their ages and of the duration of the inactive period in between. 

This represents a crucial input parameter for galaxy evolution models and simulations, which require AGN feedback to reproduce the observed galaxy mass function, as well as the correlation between the mass of the black hole and the galaxy bulge (e.g. \citealp{ferrarese2000}, \citealp{dimatteo2005}, \citealp{fabian2012}, \citealp{weinberger2017}). A comprehensive knowledge of the AGN duty cycle as a function of various source conditions, such as the black hole accretion mode, the host galaxy properties, and the surrounding intergalactic environment, is therefore vital.

Restarted radio galaxies have been known for a long time, and various authors have discussed techniques for searching and studying them. Only a few sources have been identified based on their radio spectral properties \citep{parma2007, murgia2011}. Instead, most of them have been found based on their morphology, often showing a pair of aligned double lobes (which are called `double-double radio galaxies' (DDRGs), e.g. \citealp{lara1999, schoenmakers2000, nandi2012}), or alternatively, bright compact radio jets or cores embedded in extended, low-surface brightness emission (e.g. \citealp{saripalli2012, jamrozy2007, shulevski2012, brienza2018}). Other restarted radio galaxies have individually been identified through their peculiar morphologies, for example 3C338 \citep{burns1983}, 4C~35.06 \citep{shulevski2015}, and 3C219 \citep{clarke1992}.

For a few double-double radio sources, spectral ageing modelling, based on equipartition assumptions, has provided estimates of the duration of the quiescent phase, which is found to be in the range 0.1$\sim$10 Myr. This is usually at least a factor two shorter than the first active phase, which is observed to be of the order of a few tens up to a few hundreds of million years (\citealp{konar2012, konar2013b, orru2015, nandi2019}).

\begin{figure}[!htp]
\centering
{\includegraphics[width=8cm]{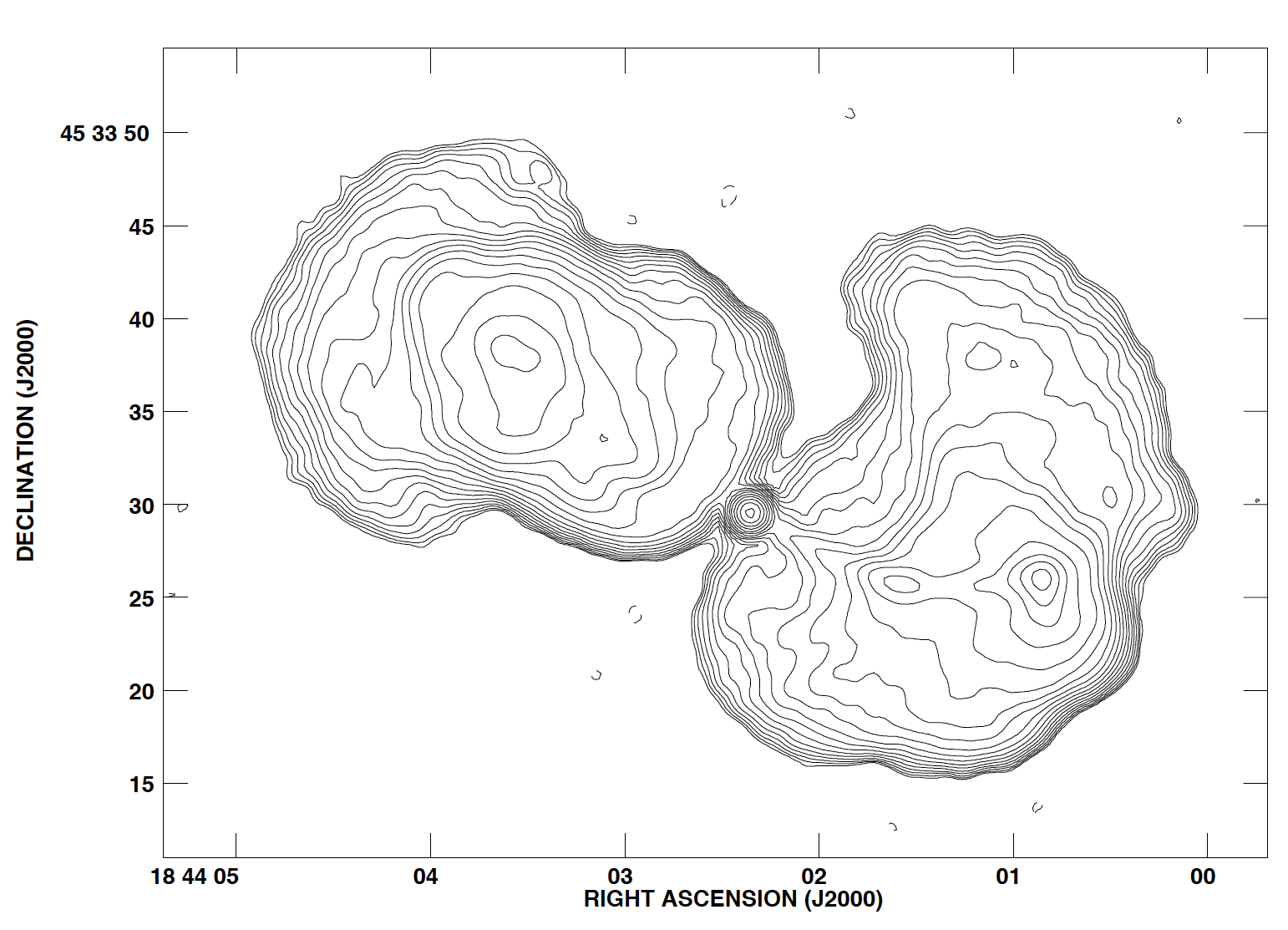}}
\caption{Radio map of 3C388 at 1.4 GHz and 1.32 arcsec resolution taken from the online atlas by \cite{leahy1996}. The lowest contour is at 0.25 mJy $\rm beam^{-1}$, and contours are separated by a ratio of $\rm \sqrt{2}$.}
\label{fig:leahymap}
\end{figure}

Major steps forward in the selection and study of restarted radio galaxies have recently been allowed by the advent of the LOw Frequency ARray (LOFAR,\citealp{vanhaarlem2013}) through its unprecedented sensitivity and resolution at low frequency, where the remnant lobes are expected to be brighter (see e.g. \citealp{shulevski2015, orru2015, brienza2018, mahatma2019, jurlin2020, shabala2020})
However, there is one peculiar class of restarted radio galaxies that we may still be missing:. sources showing the footprint of their multi-epoch activity through spatial variations of their spectral index across their radio lobes. At least one such source has been suggested in the past, the radio galaxy 3C388 (see Fig. \ref{fig:leahymap}).

3C388 is associated with a very luminous cD galaxy located in a poor 
cluster at redshift $z=0.091$ \citep{prestage1988, buttiglione2009} with a dense intracluster medium with temperature equal to 3.5 keV \citep{kraft2006, ineson2015}. The host galaxy is one of the most luminous elliptical galaxies in the local Universe ($\rm M_B=-24.24$), it shows a weak stellar nucleus in the image from the Hubble space telescope, and it is classified as a low-excitation radio galaxy \citep{jackson1997}. The radio galaxy 3C388 has an extension of about 1 arcmin, which corresponds to about 100 kpc at its redshift, and a radio luminosity equal to $\rm P_{178MHz} = 4\times10^{25} \ WHz^{-1}sr^{-1}$, lying just above the Fanaroff-Riley I/II (\citealp{fanaroff1974}) nominal border line ($\rm P_{178MHz}=2\times10^{25} \ WHz^{-1}sr^{-1}$). Its radio morphology consists of two large lobes with a broad central plateau of bright emission surrounded by extended low surface brightness emission of similar shape (see Fig. \ref{fig:leahymap}). A compact hotspot-like emission is embedded in the western lobe, well detached from the lobe edge, and is connected to the core by a narrow bent jet \citep{roettiger1994}. Based on considerations of the ratio between jet and counterjet brightness, \cite{lehay2001} estimated that the jet bends with an angle equal to $\sim$50 degrees with respect to the line of sight.

\cite{burns1982} and \cite{roettiger1994} studied the radio spectral index distribution of the radio lobes between 1.4 and 5 GHz and observed a sudden steepening towards the edges of the lobes (with maximum values of $\rm \sim\alpha_{1400MHz}^{4850MHz}$=1.6, S $\propto\nu^{-\alpha}$). Because the curvature of the radio spectrum is related to the age of the plasma, the authors suggested that the duality in the spectral index distribution indicates two different electron populations related to two different jet episodes. In particular, the authors proposed that the two reborn jets are inflating new lobes within the old remnant lobes.

An alternative interpretation proposed by \cite{burns1982} instead considers  the source to be a wide-angle tail radio galaxy as seen in projection. In this occurrence, the brighter plateau of emission would represent the region closer to the observer and would mostly be located on the plane of the sky, while the outermost low surface brightness region would belong to the tails, which develop backwards and along the observer's line of sight. While this scenario would require a very coincidental geometry and is not preferred by the authors, it is hard to completely prove it incorrect.

The new-generation instruments now enable us to obtain a resolved view of this well-known source in the MHz regime for the first time. In this paper we present a spatially resolved multi-frequency radio analysis of 3C388 aimed at further investigating its restarting nature and at constraining the timescale of the duty cycle of the jets.

This paper is organized as follows: in Sect. 2 we describe the data and the data reduction procedures; in Sect. 3 we present the source morphology and the spectral analysis; and in Sect. 4 we discuss the spectral properties of the source and the timescales of the jet activity.
The cosmology adopted in this work assumes a flat universe and the following parameters: $\rm H_{0}= 70\  km \ s^{-1}Mpc^{-1}$, $\Omega_{\Lambda}=0.7, \Omega_{M}=0.3$. At the redshift of 3C388 $z=0.091$, 1 arcsec corresponds to 1.696 kpc.

\section{Data}
\label{data}

In this section we describe the data collection and data reduction procedures that we used to perform the spectral analysis of the source 3C388. We present here new dedicated observations at 144 MHz and 350 MHz, as well as archival data at 614, 1400, and 4850 MHz. When a spectral analysis is performed, especially on resolved scales, particular attention should be paid to matching the uv-coverage of different instruments at different frequencies. This is necessary to recover the same scales of emission and avoid deriving artificial spectral trends throughout the source (see e.g. \citealp{vanbreugel1980}). For this reason, all data were chosen to have the best match in uv-coverage, with special attention to their sensitivity to the large scale structures. The largest angular scale that can be observed by an interferometer is 0.6$ \lambda/ D_{min}$, where $D_{min}$ is its shortest baseline (Tamhane et al. 2015). Using this expression as a reference, we ensured that all the observations we used were sensitive to angular scales $\ge$1 arcmin, which is the angular size of 3C388. 
A summary of the radio observations and image properties is presented in Table~\ref{tab:data}. 
In addition to the radio data, we also present archival Chandra observations, which we used to search for X-ray inverse-Compton emission from the radio lobes to constrain their magnetic field strength.

\begin{table*}[t]

        \small
 \caption{Summary of the radio observation and image properties. All images have a restored beam of 6 arcsec $\times$ 6 arcsec. An asterisk indicates archival observations.}
        \centering
                \begin{tabular}{c c c c c c c c c}
                \hline
                \hline
                Telescope & Configuration & LAS$^1$ & Frequency & Bandwdith & Target TOS$^2$ & Calibrators & Observation date  & Noise \\
                && [arcmin] &  [MHz] & [MHz] & [hr] &  & & [mJy beam$^{-1}$] \\
                \hline
                \hline
                
                LOFAR& HBA Inner & 244& 144 & 48 & 8 & 3C295 &  26 June 2019  & 0.8\\

                VLA & A  & 2.5 & 350 & 256& $\sim$2 & 3C286  & 28 July 2015 & 1.5\\

                GMRT & - & 17 & 614 & 16 & $\sim$4 &  3C48, 1829+487 & 29-30 May 2005* & 0.8\\

                VLA & B & 4 & 1400 & 50 & 7 & 3C286, 1843+400  & August 1986* & 0.5\\
                VLA & C & 3.5& 4850 & 50 & $\sim$5 & 3C286, 1843+400  & December 1986*  & 0.1 \\
                \hline
                \hline  
                \end{tabular}
                
                \begin{tablenotes}
                \item $^1$ Largest Angular Scale; $^2$ Time On Source.
                \end{tablenotes}
     \label{tab:data}
\end{table*}

\subsection{LOFAR observations at 144 MHz and data reduction}

We performed a targeted observation of the source 3C388 with the LOFAR High Band Antennas (HBA, 150 MHz) on March 2, 2014, in the framework of the Surveys Key Science Project (LC1\_034). However, the quality of this dataset did not allow us to reach a good enough sensitivity and image fidelity, even using the most advanced data reduction pipelines available to date.

For this work we therefore used a more recent dataset obtained on June 26, 2019, as part of the LOFAR Two-metre Sky Survey (LoTSS, see \citealp{shimwell2019}). The pointing reference code is P280+45 (project LT10\_010), and the target lies at a distance of 0.36 degrees from the field phase centre. The observations were carried out using the standard survey setup, with 8 hours on-source time, 48 MHz bandwidth (244 subbands), and 1 second integration time. The entire LOFAR array was used in the observations (Dutch and international stations), but for this work, we only exploit the data collected by the Dutch array, which consists of 64 stations and provides a maximum baseline of $\sim$100 km. The source 3C295 was used as flux density calibrator and was observed for 10 minutes before and after the target observation. A full description of the observing strategy of the LoTSS pointings can be found in \cite{shimwell2019}.

The data were first pre-processed using the observatory pipeline \citep{heald2010}, which includes automatic flagging of radio frequency interference (RFI) using the AOFlagger \citep{offringa2012} and data averaging in time and frequency down to 5 seconds per sample and four channels per sub-band. Aftewards, the calibration scheme and pipelines developed for LoTSS were applied (see \citealp{shimwell2019} for more details). In particular, the PREFACTOR pipeline \citep{vanweeren2016, williams2016} and version v2.2-167 of the DDFacet pipeline\footnote{https://github.com/mhardcastle/ddf-pipeline} (see \citealp{tasse2018}; Tasse et al. in prep.) were used to perform direction-independent and dependent calibration, respectively, using the default parameters. The flux scale was set according to \cite{scaife2012}.

To further improve the image quality of the target, after running the pipeline, we subtracted from the uv-data all the sources located in the field of view other than 3C388 and a few neighbour sources, and performed additional phase and amplitude self-calibration loops (van Weeren et al., in prep.). We then re-imaged the target using WSClean version 2.7 \citep{offringa2014} with a uniform weighting scheme and a restoring beam of 6 arcsec $\times$ 6 arcsec. This final image has an average RMS of 0.8 mJy $\rm beam^{-1}$, which increases up to $\sim$5 mJy $\rm beam^{-1}$ close to the target due to dynamic range limitations.
The new radio map of the source is presented in Fig. \ref{fig:pband_image} (top panel).

\subsection{VLA observations and data reduction}

\subsubsection{P band}

We observed the source with the Very Large Array (VLA) in A configuration on July 28, 2015, using the P-band receiver centred at 350 MHz. The target was observed for 2 hours, and the flux density calibrator, 3C286, was observed for 10 minutes at the beginning of the observing run. We used a correlator integration time of 2 seconds and recorded four polarization products (RR, LL, RL, and LR). The total bandwidth, equal to 256 MHz in the range 224-480 MHz, was divided by default into 16 sub-bands of 16 MHz with 128 frequency channels. 

The data were calibrated and imaged using the Common Astronomy Software Applications (CASA, version 5.1.1-5, \citealp{mcmullin2007}) in the standard manner and following the guidelines set out in the online guidelines for continuum P-band data\footnote{\url{https://casaguides.nrao.edu/index.php/VLA_Radio_galaxy_3C_129:_P-band_continuum_tutorial-CASA4.7.0}}. The flux scale was set according to \cite{scaife2012}. Nine sub-bands spread across the band were discarded due to severe RFI contamination.  

The remaining seven good sub-bands were imaged together using multiscale CLEAN with scales [0, 5, 15, 45] and nterms=2. This image was used as the starting model to perform phase self-calibration on each sub-band independently. To reduce the computational time during imaging, each sub-band was averaged down in frequency to 16 channels of 1 MHz bandwidth, but no averaging in time was performed. The final image for each subband was made using Briggs weighting with a robust parameter of 0.0. The images were restored with a beam of 6~arcsec~$\times$~6~arcsec and have noise equal to $\rm \sim 1.5 \ mJy \ beam^{-1}$. The final image of the spectral window centred at 392 MHz is presented in Fig. \ref{fig:pband_image} (bottom panel) for illustration purposes.

\subsubsection{L band and C band}

We reprocessed the data used by \cite{roettiger1994} at 1400 MHz and 4850 MHz. The data consist of observations in  B array at 1400 MHz and in C array at 4850 MHz. The target was observed for 7 hours at 1400 MHz and for $\text{about }$5 hours at 4850 MHz. The source 3C286 and 1843+400 were used as flux density calibrator and phase calibrator, respectively. The correlator integration time was set to 10 seconds.

All datasets were reduced with the standard approach using CASA (version 4.7). The data were manually flagged and calibrated using the flux scale of \cite{perley2013}, which is consistent with the scale of \cite{scaife2012} at low frequency. Phase and amplitude self-calibration were performed. The final images were obtained using Briggs weighting with robust  parameter equal to 0.0 and multiscale option with scales [0, 5, 15, 45]. The images were restored with a beam of 6~arcsec~$\times$~6~arcsec and have a noise equal to 0.5 $\rm mJy \ beam^{-1}$ at 1400 MHz and equal to 0.1 $\rm mJy \ beam^{-1}$  at 4850 MHz.

\subsection{GMRT observations at 614 MHz and data reduction}

The target was observed with the legacy Giant Metrewave Radio Telescope (GMRT) at 614 MHz and 240 MHz in dual-frequency mode, and data were published in \cite{lal2008}. The observations were performed on July 29 and 30, 2005. The target 
observation was divided into five time-scans for a total integration time of 4 hours. The source 3C48 was used as flux density calibrator and 
observed for 10 minutes at the beginning and  end of the observing 
session. Data were recorded using a correlator integration time of 16.1 
seconds and a total bandwidth of 33 MHz divided into 512 channels of 65 kHz each.

For this work we reprocessed the archival data at 614 MHz using the SPAM pipeline \citep{intema2014,intema2017} and set the absolute flux scale according to \cite{scaife2012}. The output calibrated visibility data were imported into CASA to produce images at different resolutions. The final image was obtained using a Briggs weighting scheme with a robust parameter equal to 0.0 and a restoring beam equal to 6~arcsec~$\times$~6~arcsec. The noise in the final image is equal to $\rm \sim 0.8 \ mJy \ beam^{-1}$.
The available dataset at 240 MHz was not included in this analysis because its spatial resolution is lower than in the other datasets.

\subsection{Chandra observations and data reduction}
\label{chandra}

3C388 was observed by Chandra on February 9 and 29, 2004, with the ACIS-I detector (obs ID 4756 and 5295, respectively), and the data were published by \cite{kraft2006}. We reprocessed the archival observations using the Chandra Interactive Analysis of Observations CIAO software package \citep{fruscione2006} from the level 1 events files with \textsc{CIAO} 4.8 and CALDB 4.7.0. 
The \textit{chandra\_repro} pipeline was subsequently used to reprocess the data to produce new level 2 event files using standard CIAO analysis methods. The \textsc{reproject\_obs} tool was used to merge the event files from the two observations, resulting in the 0.5-7.0 keV image shown in Fig. \ref{fig:Xspectrum} (top panel).

\begin{figure}[!htp]
\centering
{\includegraphics[width=8cm]{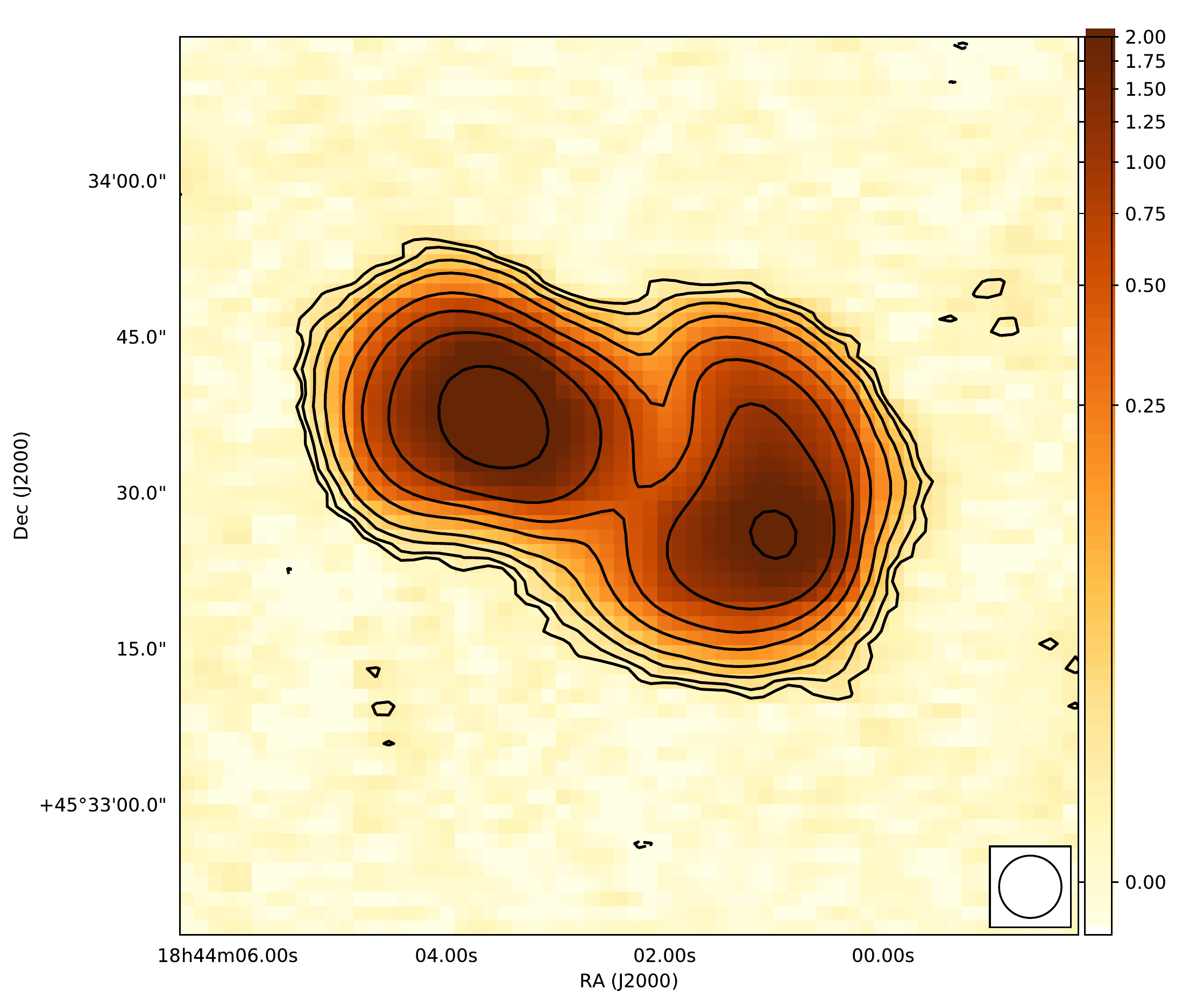}}
\centering
{\includegraphics[width=8cm]{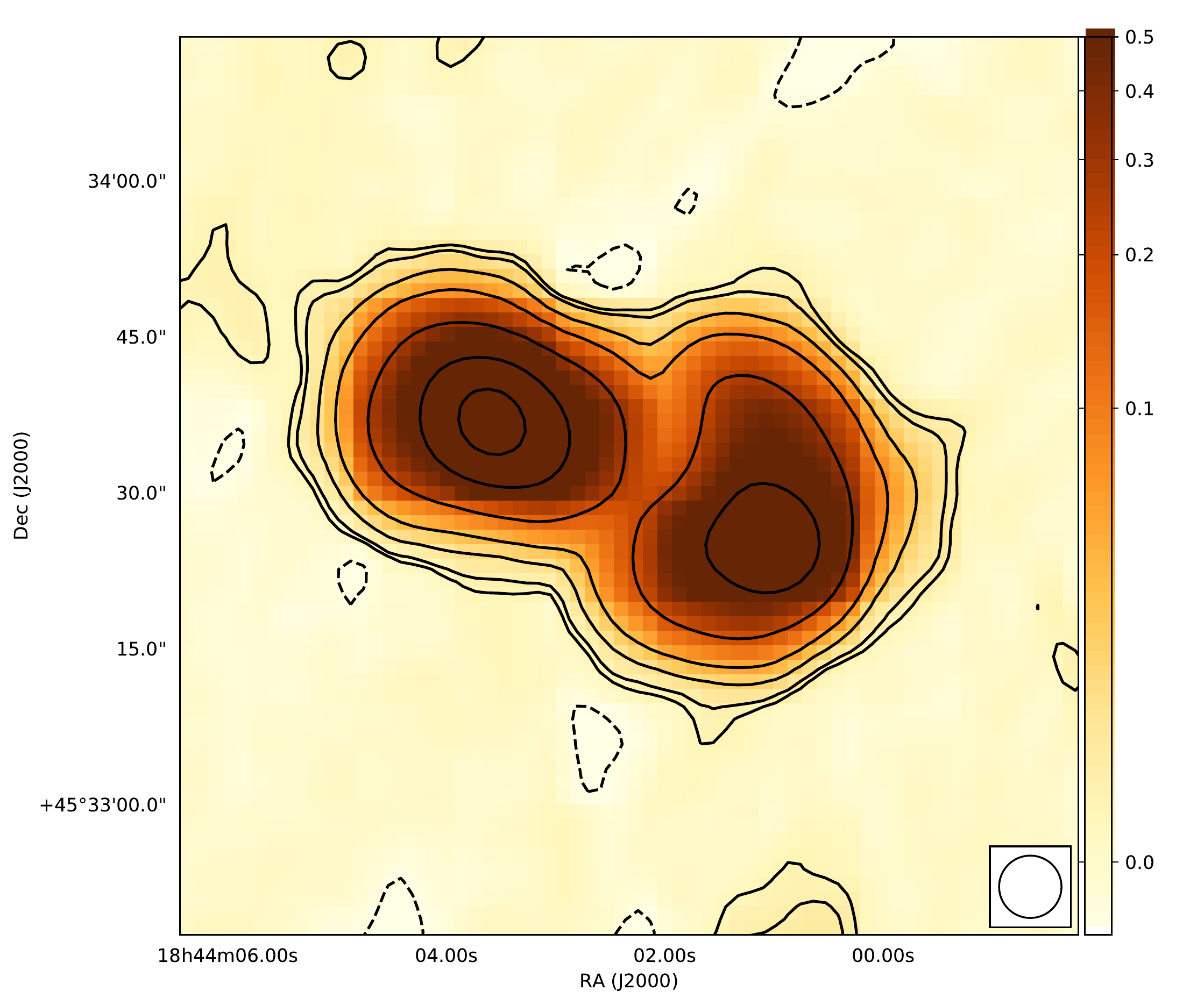}}
\caption{New low-frequency radio maps of 3C388, which show that the morphology of the source is consistent with previous observations at GHz frequencies. $Top$  LOFAR 144-MHz map with contours equal to -3, 3, 5, 10, 20, 50, 100, 200, and 500 $\times  \ \sigma_{144}$ , where $\sigma_{144}$=5 mJy $\rm beam^{-1}$. $Bottom$ VLA 392-MHz map with contours equal to -3, 3, 5, 15, 40, 150, 500, and 1000 $\times  \ \sigma_{392}$ , where $\sigma_{392}$=1.5 mJy $\rm beam^{-1}$. Colour-bars are expressed in Jy $\rm beam^{-1}$.  }
\label{fig:pband_image}
 \end{figure}

\section{Analysis}
\label{results}

\subsection{Morphology}

In Fig.~\ref{fig:pband_image} we show the new low-frequency radio maps of the source at 144 and 392 MHz. The morphology of the source at low frequency agrees with previous observations at higher frequency \citep{burns1982, roettiger1994, leahy1996, lal2008}. Using both the LOFAR and the P-band VLA image, we measure an angular size of $\sim$1 arcmin, using the $\rm 5\sigma$ contours as a reference, which corresponds to $\sim$100 kpc at $z=0.091$. This is consistent with previous estimates at higher frequencies. 

Interestingly, in the LOFAR map we do not detect any additional low surface brightness emission beyond the known shape of the lobes at higher frequency. This might have been expected if the source were a wide-angle tail seen in projection, as described in Sect.~\ref{intro}. Tailed radio galaxies in LOFAR images indeed often appear much more extended and asymmetric than what is observed at higher frequency. The most striking example to date is NGC 326 (\citealp{hardcastle2019}). While limited in dynamic range, our current image therefore appears to suggest that the plasma of the lobes of 3C388 is well confined.
Finally, we note that the maximum resolution imposed by the low frequency (equal to 6~arcsec~$\times$~6~arcsec) does not allow us to investigate the small-scale structures recognised in previous studies, such as the narrow jet and hotspot in the western lobe \citep{roettiger1994}.

\subsection{Integrated radio spectrum}
\label{spec}

\begin{figure}
\centering
{\includegraphics[width=7.5cm]{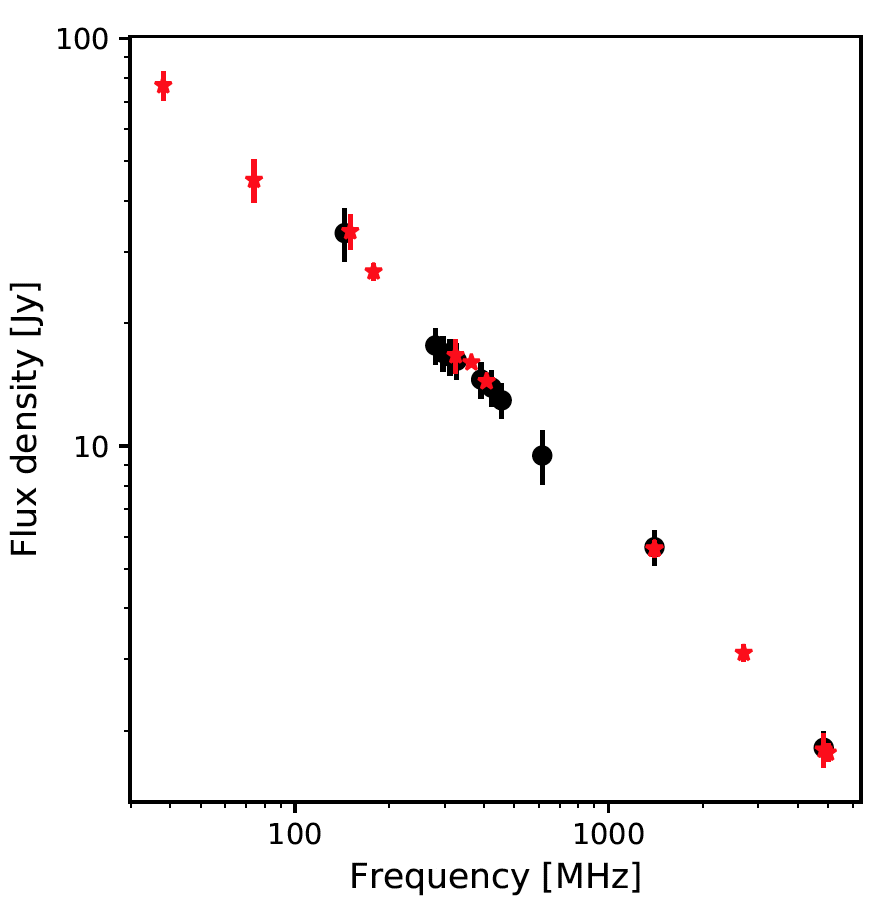}}
\caption{Integrated radio spectrum of the source 3C388, which shows that the measurements presented in this work (black circles) agree well with the spectral behaviour known from the literature (red stars). All measurements taken from the literature have been set to same flux scale used in this work (see Sect. \ref{data}). The list of all flux densities is presented in Table~\ref{tab:fluxes}.}
\label{fig:int_spectrum}
 \end{figure}

In Fig. \ref{fig:int_spectrum} we show the integrated radio spectrum of 3C388 that we reconstructed to verify the quality of the flux calibration of our different observations with respect to previous works. The measurements presented in this paper (shown as black circles in the plot) agree weel with the overall spectral shape known in the literature (shown with red stars in the plot). The integrated flux density at each frequency was measured using the $\rm 5\sigma$ contours as a reference. In Table \ref{tab:fluxes} we show the list of flux density measurements with the respective errors. We note that all measurements taken from the literature were set to same flux scale used in Sect. \ref{data}. The errors on the flux densities were computed by combining in quadrature the flux scale error and the image noise, as shown in \cite{klein2003}. However, the main source of uncertainty is related to the flux scale. In particular, the flux scale error is considered to be 15\%\ for LOFAR \citep{shimwell2019} and GMRT measurements, 5\%\ for VLA measurements at P band, and 2\%\ at L and C band \citep{scaife2012, perley2013}. 

As extensively described by \cite{shimwell2019}, the LOFAR flux scale may suffer from severe systematic offsets, therefore we paid this particular attention. Figure \ref{fig:int_spectrum} shows that the LOFAR measurement at 144 MHz is well aligned with the 150 MHz all-sky radio survey (TGSS ADR1, \citealp{intema2017}). While both the LOFAR and TGSS measurements appear to be slightly upscaled with respect to the general spectral trend, they are consistent with the general spectral shape within the applied errors, therefore we did not perform any additional flux scaling.

\begin{table}[htp!]
\small
\caption{Integrated flux densities of 3C388 at various frequencies measured in this work and presented in the literature. All measurements taken from the literature have been set to same flux scale used in this work (see Sect. \ref{data}). }
        \centering
                \begin{tabular}{r r l}
                \hline
                \hline
                Frequency & Flux density  & Reference \\
                $\rm [MHz]$ & [Jy] & \\
                \hline
                \hline
                38    &  76.70$\pm$6.40    &  \cite{kellerman1969}\\
                74        &  44.97$\pm$5.46     & \cite{lane2012} (VLSSr)\\
                144   &  33.30$\pm$5.00 &  this work\\
                150   & 33.62$\pm$3.36 & \cite{intema2017} (TGSS)\\
                178   &  26.80$\pm$1.34 &  \cite{kellerman1969}\\
                280   &  17.65$\pm$0.88 &  this work\\
                296   &  16.93$\pm$0.85 &  this work\\
            312   &  16.59$\pm$0.83 &  this work\\
            325   &  16.70$\pm$1.60 &   \cite{rengelink1997} (WENSS)\\
            328   &  16.20$\pm$0.81 &  this work\\
            365   &  16.04$\pm$0.40 &   \cite{douglas1996} (TXS)\\
            392   &  14.56$\pm$0.73 &  this work\\
            408   &  14.42$\pm$0.29 &   \cite{ficarra1985} (B3.1)\\
            424   &  13.91$\pm$0.69 &  this work\\
            456   &  12.95$\pm$0.64 &  this work\\
            614   &  9.48$\pm$1.42  &  this work\\
            1400  &  5.60$\pm$0.28   &   \cite{kellerman1969}\\
            1400  &  5.60$\pm$0.19              & \cite{condon1998} (NVSS)\\
            1400  &  5.65$\pm$0.11  &   this work\\
            2696  &       3.11$\pm$0.15 &   \cite{kellerman1969}\\
            4850  &  1.82$\pm$0.04  &  this work\\
            4850  &  1.80$\pm$0.17 & \cite{gregory1996} (GB6) \\
            5000  &      1.77$\pm$0.09  &   \cite{kellerman1969}\\      
            \hline                  
                \hline                  
                \end{tabular}          
        \label{tab:fluxes}          
\end{table}                     

\subsection{Magnetic field}
\label{magn}

Because it regulates the amount of radiative losses in the lobes of radio galaxies at low redshift (where the inverse-Compton scattering with the cosmic microwave background, CMB,  is not dominant), the magnetic field represents the most crucial input parameter of spectral ageing models. We discuss these models in Sect.~\ref{pixelage}.

A direct estimate of the magnetic field strength, as well as of the number density of the emitting particle in the lobes, can be obtained when the lobes are detected at radio and X-ray frequencies at the same time.  X-ray emission in the lobes of radio galaxies is indeed thought to originate from inverse-Compton scattering between the same relativistic electrons that produce the observed radio synchrotron radiation and the CMB \citep{harris1979}.

When no X-ray observations are available, it is common practice to rely on the equipartition assumption for the magnetic field calculation. However, in recent years, an increasing number of studies of samples of FRII radio galaxies have demonstrated that magnetic field strengths in these sources are typically a factor of 2 - 3 below the equipartition values (e.g. \citealp{croston2005}, \citealp{kataoka2005}, \citealp{migliori2007} and \citealp{ineson2017}, \citealp{turner2018}). 

In order to assess the magnetic field strength of the source 3C388 in the best possible way, we computed the equipartition value $\rm B_{eq}$ and attempted to derive the magnetic field $\rm B_{IC}$ from the X-ray inverse-Compton emission as described below. In the remainder of the paper we consider these two magnetic field values when we perform the spectral modelling to explore the resulting variations in the source age.

\subsubsection{Inverse-Compton constraints to the magnetic field}
\label{magn-ic}

\begin{figure}[htp!]
\centering
{\includegraphics[width=8.5cm]{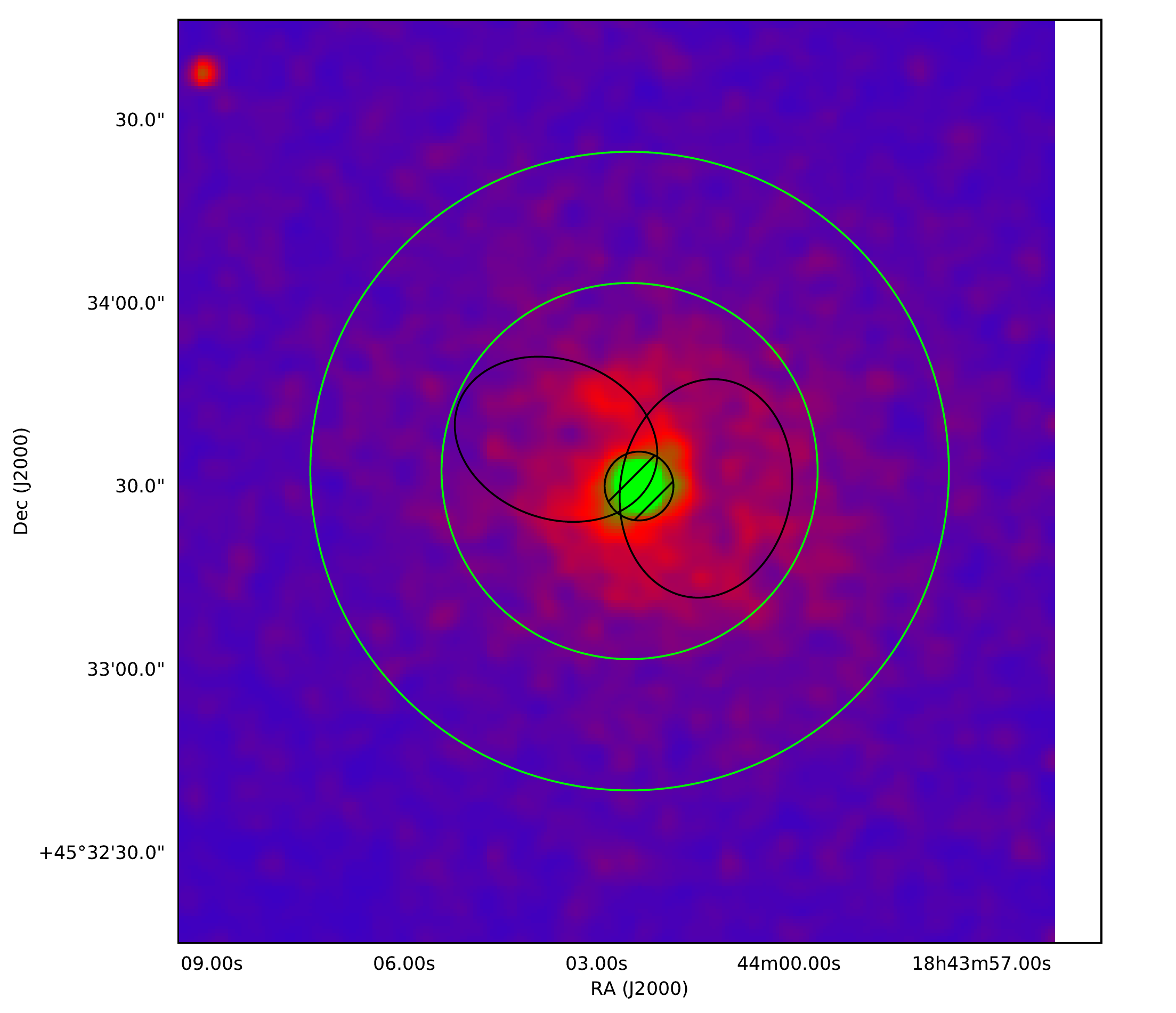}}
{\includegraphics[width=8.5cm]{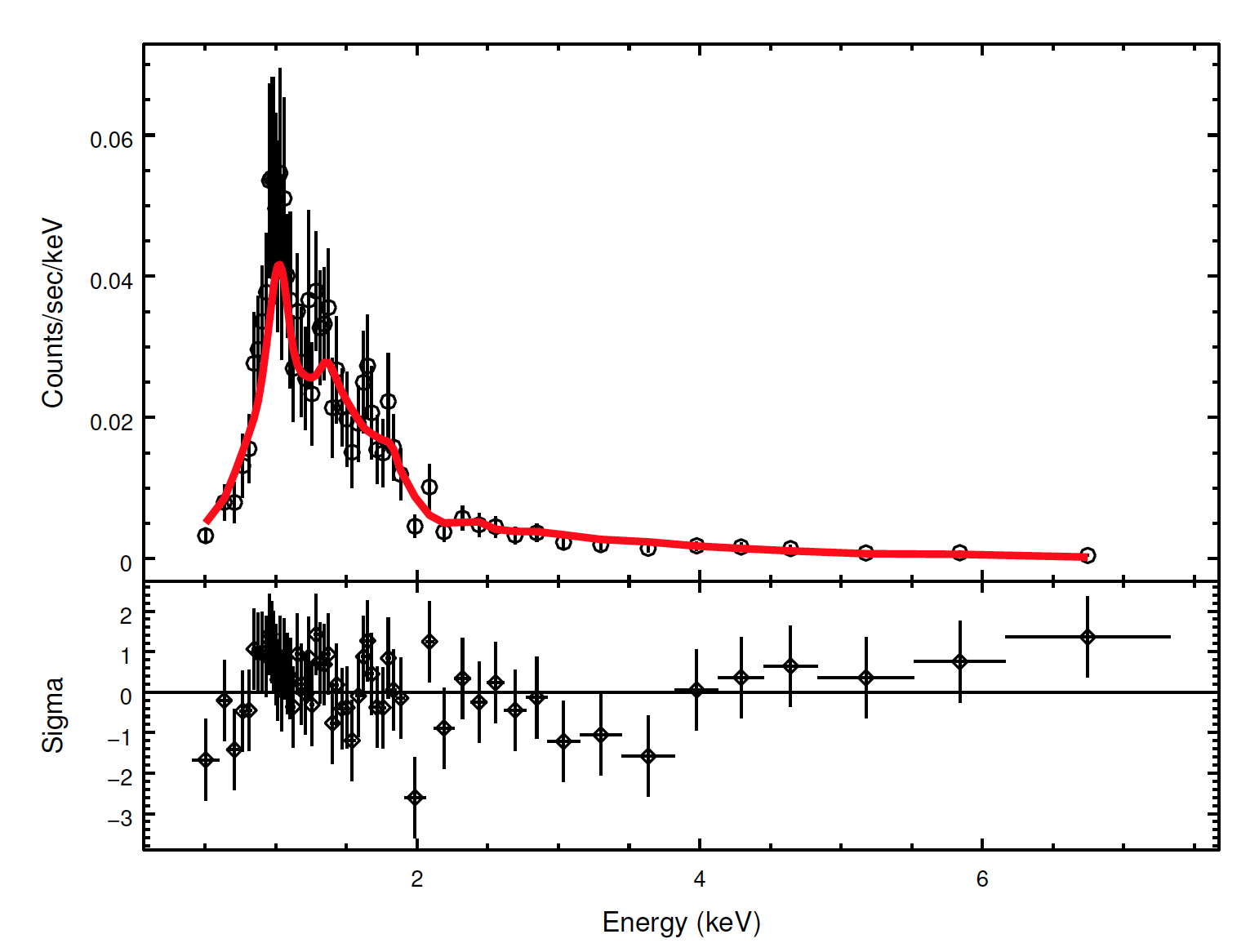}}
\caption{\textit{Top}: Chandra X-ray map of 3C388 (0.5-7 keV). The green annulus has been used for background subtraction, while the black regions have been used to extract the source spectrum. The central circle with a diagonal line shows the central AGN core, which was masked out from the analysis. \textit{Bottom}: Chandra background-subtracted X-ray spectrum for 3C388 extracted from the black regions shown in the top panel. The red solid line indicates the best fit to the data as described in Sect. \ref{magn-ic}, while the lower panel indicates the reduced $\rm \chi^2$ statistics at each data point.}
\label{fig:Xspectrum}
 \end{figure}

In order to obtain a measurement of the inverse-Compton emission in the lobes of 3C388 and a constraint on the magnetic field strength, we used the Chandra X-ray data described in Sect. \ref{chandra} following the strategy presented below.
We chose to use a circular annulus surrounding the radio lobes as a reliable region to estimate the required background subtraction to remove foreground cluster emission in front of the lobes. We also removed the central AGN component that might contaminate the desired lobe spectrum. We extracted count rates from the individual observations, which were then combined using CIAO tools. 
The background regions and the lobe regions we used to extract the spectrum are shown in Fig. \ref{fig:Xspectrum} (top panel).

We then used the Sherpa application (Freeman et al. 2011) as a platform to load the spectra and to perform background subtraction and model fitting. Bad data were removed using the \textit{ignore\_bad()} command, which removes bins based on bad data flags.

We fitted the lobe spectrum using a combination of a non-thermal power-law emission model, describing the inverse-Compton emission, and a thermal model (APEC), describing collisionally ionised diffuse gas, as is predominantly expected from X-ray emission of the ICM. 
A photo-electric absorption model was also added as a multiplicative component to the APEC model to account for the absorption of X-ray photons by foreground atomic matter such as hydrogen atoms in cold gas. 
The Galactic absorbing column density was fixed at $6.32 \times 10^{20}$~atoms~cm$^{-2}$ \citep{dickey1990}. The temperature of the APEC model was set to the X-ray temperature in the lobe regions equal to 3.5 keV, as measured by \cite{kraft2006}. We note that indications of a temperature variation in different regions of the source have been presented by \cite{kraft2006}. These variations are not very significant, however, because the measured temperatures are affected by large uncertainties. For this reason, we prefer to assume a single-temperature average value throughout the source.

Subsequent fitting resulted in the power-law photon index $\Gamma$ being unconstrained, and we therefore fixed this value at $\Gamma=1.57$ because our best-fit injection spectral index is equal to $\alpha_{inj}$ = 0.57 (see Sect. \ref{pixelage} for a full discussion) and $\Gamma = \alpha_{inj} + 1$. The use of the injection index is justified because the electrons that scatter the CMB to X-ray energies are those with the lowest energies ($\gamma\sim$1000). The re-fitting of the model with only the power-law and thermal normalization as free parameters is consistent with a non-detection of non-thermal power-law emission, with a reduced $\chi^2 = 0.8$. The best fit is shown in Fig. \ref{fig:Xspectrum} (bottom panel). A 3$\sigma$ upper limit flux density on the inverse-Compton emission is found at 1 keV equal to $0.0102 \ \mu$Jy.

We then used the \textsc{SYNCH} code \citep{hardcastle1998} to determine the magnetic field strength that simultaneously matched the observed radio flux densities reported in Table \ref{tab:fluxes} (including those from literature) and the upper limit of the X-ray flux density derived from the Chandra image. Using a model that assumes no proton content in the lobes, we found a lower limit on the magnetic field strength of $\rm B_{IC}> 3 \ \mu$G.

\subsubsection{Equipartition magnetic field}
\label{magn-eq}

Because the magnetic field estimate based on the inverse-Compton emission $\rm B_{IC}$ described in Sect. \ref{magn-ic} only provided us with a lower limit, we also derived the equipartition magnetic field value equal to $\rm B_{eq}=15.8 \ \mu G$ using the derivation by \cite{worrall2006}. This relies on the assumption of equipartition conditions between particles and magnetic field over the entire source. For the calculation we assumed a power-law particle distribution of the form $N(\gamma)\propto\gamma^{-p}$ between a minimum and maximum Lorentz factor of $\gamma_{min}=10$ and $\gamma_{max}$=$10^6$, with $p$ being the particle energy power index. The value of $p$ relates to the injection spectral index $\alpha_{inj}$ of the synchrotron power spectrum (S $\propto \nu^{-\alpha_{inj}}$) as $p=2\alpha_{inj}+1$ and therefore was set to $p=2.1$ following the best-fit value equal to $\alpha_{inj}$=0.57 (see Sect. \ref{pixelage} for a full discussion). The ratio between proton and electron content inside the lobes was assumed to be $k=U_{p}/U_{e}=$ 0, as has been suggested to be the case in many FRII radio galaxies (e.g. \citealp{croston2005}). To calculate the volume of the source, we assumed the two lobes to be ellipsoids with major axes equal to a=34 arcsec and a=36 arcsec and minor axes equal to b=26 arcsec and b=28 arcsec for the western and eastern lobe, respectively. A value of $S_{1400}=5.6$ Jy was used as a reference. 

We note that the lower limit of the magnetic field obtained from the X-ray data equal to $\rm B_{IC}$ > 3 $\rm \mu$G is a factor $\sim$5 lower than the equipartition value. This is consistent with what has been observed in many sources, as discussed in Sect. \ref{magn}, suggesting that the real magnetic field value of 3C388 lies in the range 3$\sim$15.8~$\rm \mu$G and is likely closer to the computed $\rm B_{IC}$.

\subsection{Spatially resolved spectral analysis}
\label{resolved}

The shape of the radio spectrum of jetted AGN can give interesting insights into the physics of the electron population that causes the emission. To investigate the spectral shape of the plasma in different regions of the lobes, we performed a spatially resolved analysis as described below. We imaged all the data using the same pixel size equal to 1.2 arcsec and a final restoring beam equal to 6~arcsec~$\times$~6~arcsec. Moreover, we spatially aligned the maps to correct for any possible spatial shifts introduced by the imaging and phase self-calibration process, which would compromise the reliability of the spectral analysis. To do this, we fitted a point source located near the target with a 2D Gaussian function in all the available images and derived the central pixel position. We then used one image as a reference and aligned all the others with it, using the tasks IMHEAD and IMREGRID in CASA. After this procedure, we find a maximum residual offset between the images $\rm \leq$0.1 pixels, which is sufficiently accurate for our analysis.

For the spectral analysis, we used the broadband radio astronomy tools software package\footnote{http://www.askanastronomer.co.uk/brats/} (\texttt{BRATS}, \citealp{harwood2013, harwood2015}). Here we describe the procedures used in this analysis, and we refer to the cookbook for a complete description of the methods underlying the software.

\subsubsection{Spectral index maps}
\label{sindex}

\begin{figure*}[htp!]
\minipage{0.33\textwidth}
  \includegraphics[width=1\textwidth]{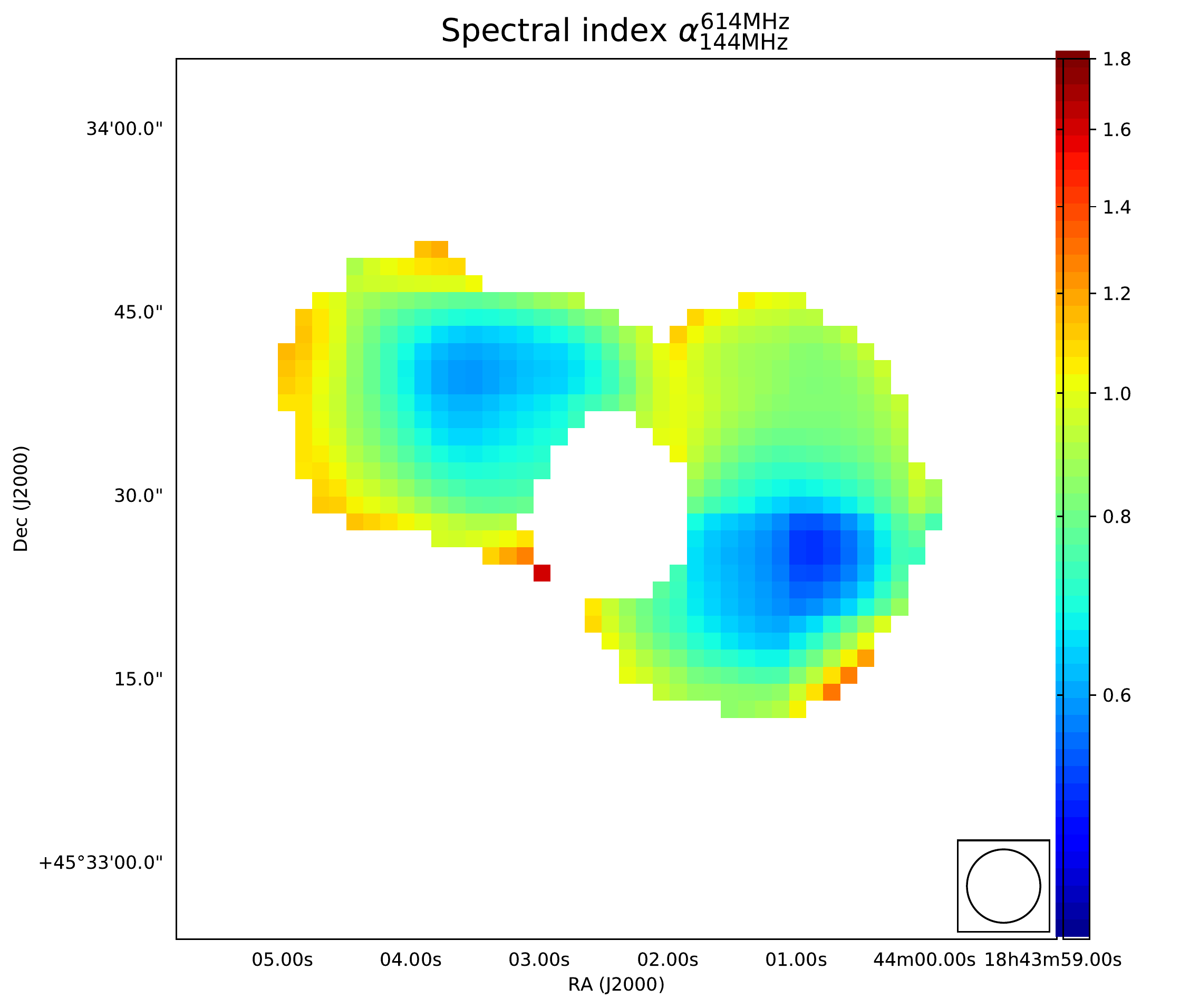}

\endminipage\hfill
\minipage{0.33\textwidth}
  \includegraphics[width=1\textwidth]{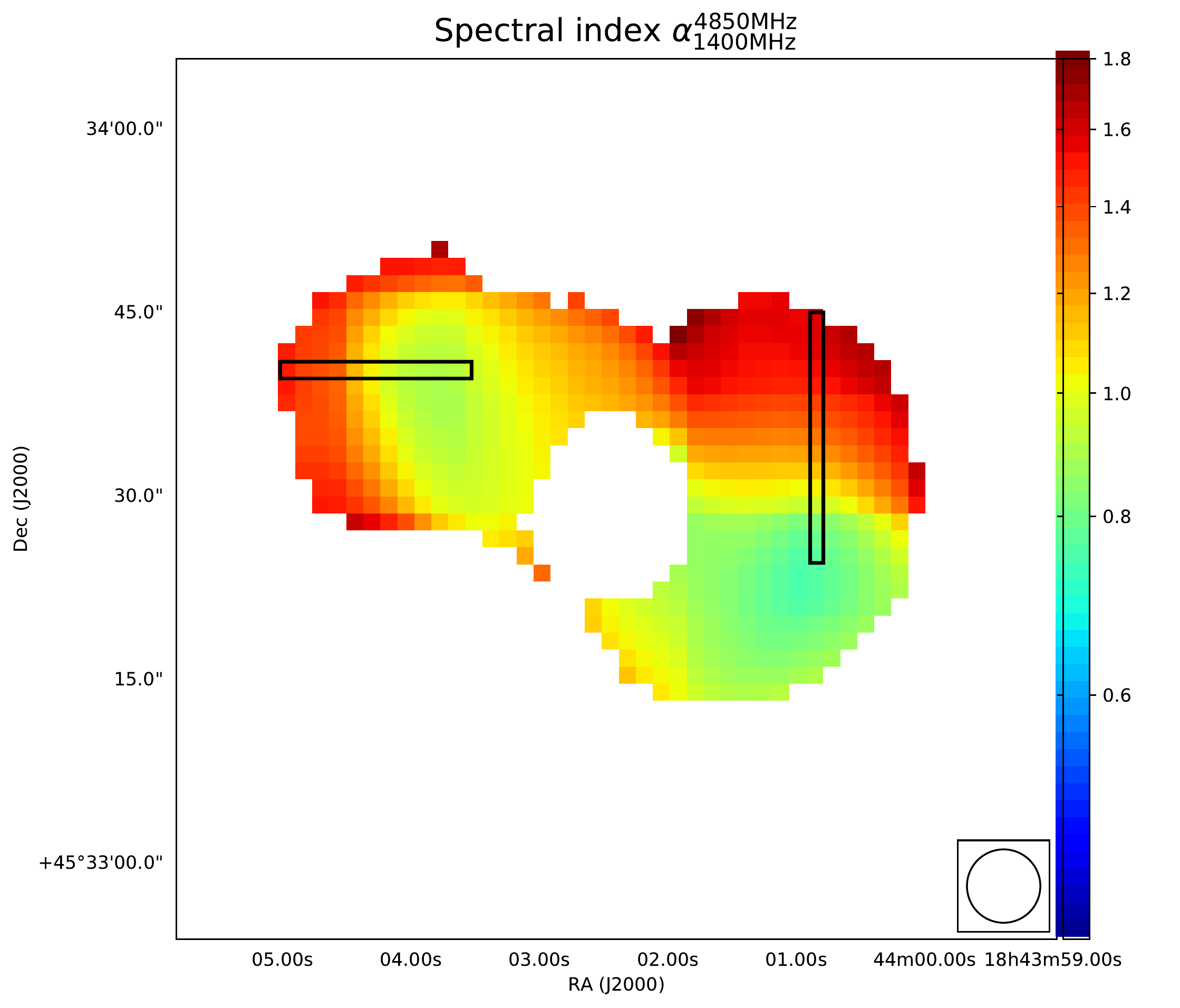}

\endminipage\hfill
\minipage{0.33\textwidth}
  \includegraphics[width=1\textwidth]{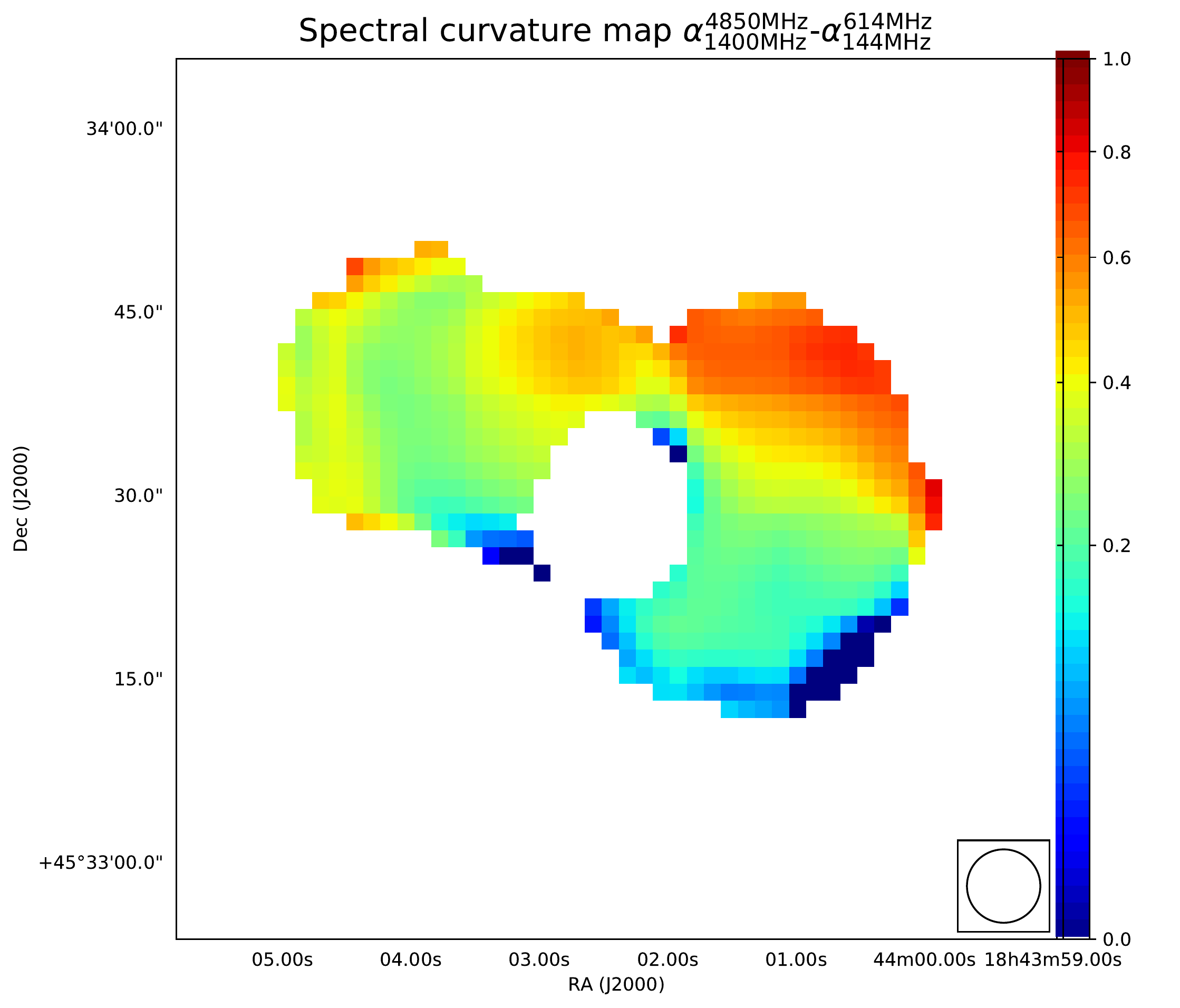}
 \endminipage\hfill 
 
\caption{$Left$: Spectral index map in the range 144-614 MHz. $Middle$: Spectral index map in the range 1400-4850 MHz. $Right$: Spectral curvature map SPC = $\alpha_{\rm 1400 MHz}^{\rm 4850 MHz}$ - $\alpha_{\rm 144 MHz}^{\rm 614 MHz}$. All three maps show a clear steepening moving from the inner regions of the lobes towards the edges of the lobes. The spatial resolution of all radio maps is 6~arcsec~$\times$~6~arcsec. The black rectangular regions in the middle panel have been used to study the spectral index variations throughout the lobes (see Fig. \ref{fig:spec_map_distr}).}
\label{fig:spec_map}
\end{figure*}

We computed two spectral index maps at low and high frequency, respectively, using the weighted least-squares method and only considering pixels above 5$\sigma$ in each single-frequency map. For the low-frequency spectral index map we used all the images in the range 144-614 MHz, while for the high-frequency spectral index map we used the images at 1400 and 4850 MHz. We note that we excluded the P-band image centred at 280 MHz (see Table \ref{tab:fluxes}) from the following analysis because its image fidelity is poor. Our final goal is to analyse the spectral behaviour of the lobes, therefore we excluded the region corresponding to the core of the radio galaxy from the analysis. Moreover, the core is not well described by any of the spectral ageing models presented in Sect. \ref{pixelage}. The spectral index maps are shown in the left and middle panels of Fig. \ref{fig:spec_map}. The errors in the low-frequency spectral index map vary from 0.1 up to 0.2 in the most external regions, and in the high-frequency spectral index map they vary from 0.01 up to 0.05 in the most external regions.

\begin{figure}[htp!]
\center
\includegraphics[width=0.45\textwidth]{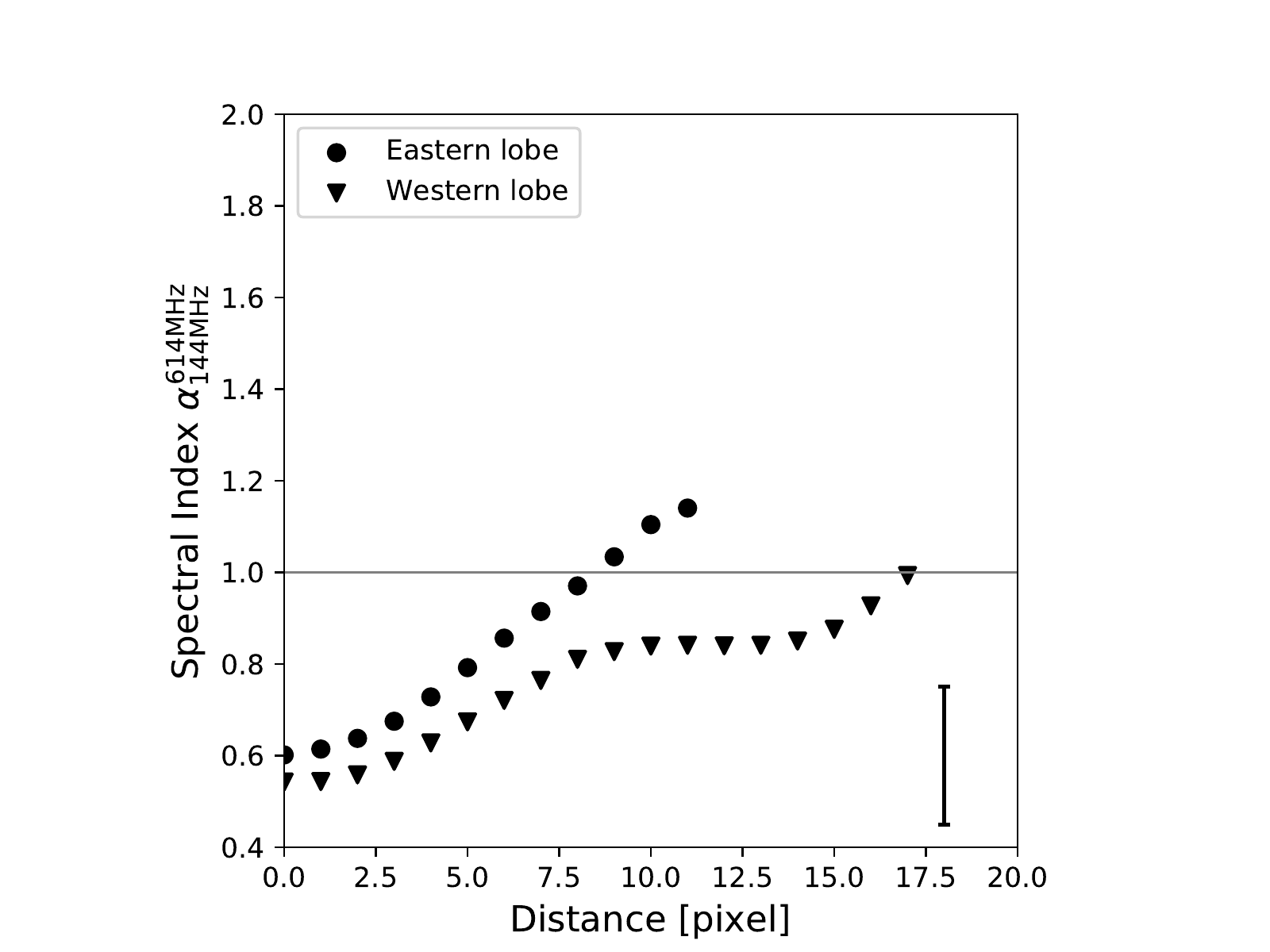}
\includegraphics[width=0.45\textwidth]{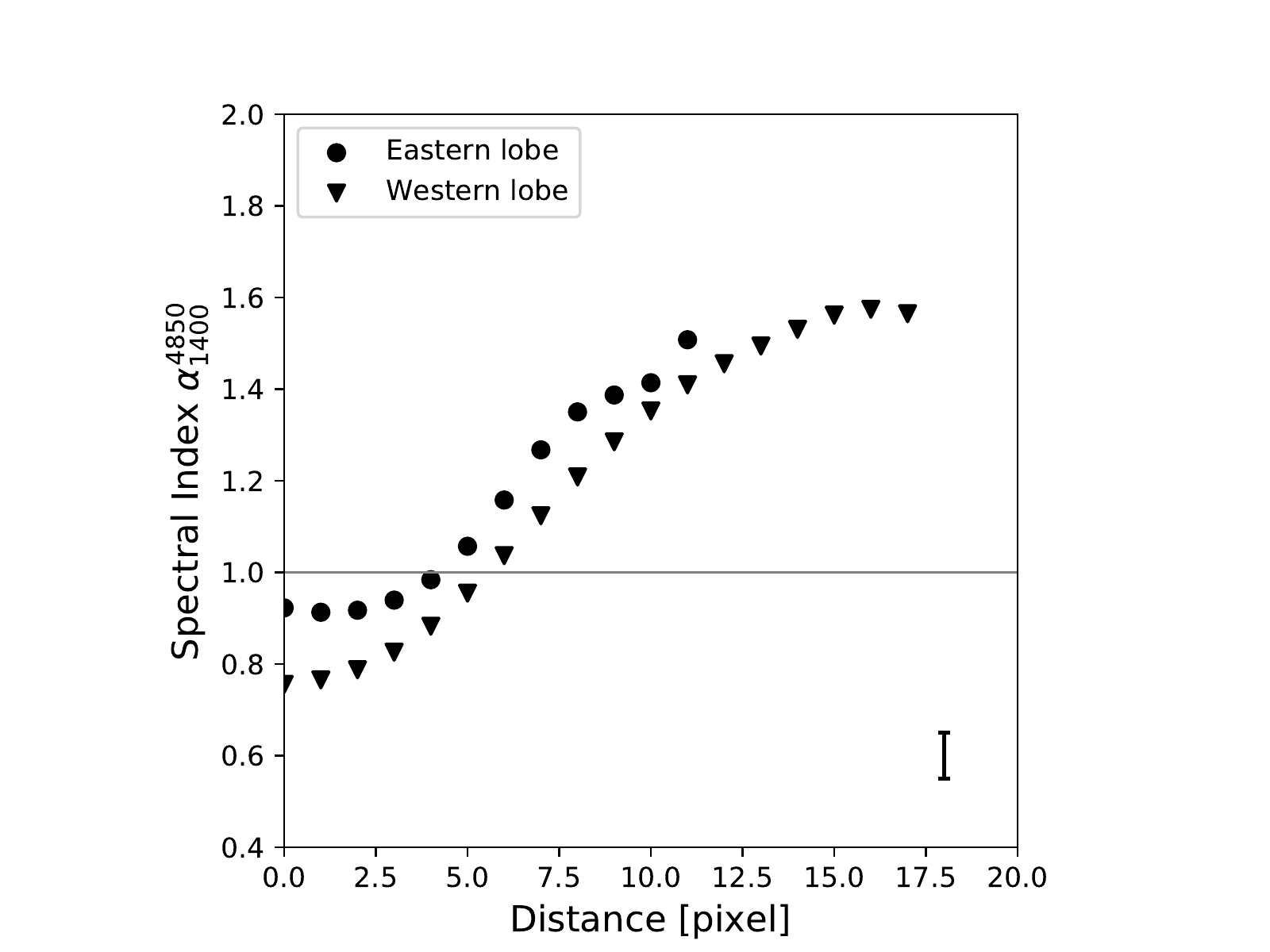}

\caption{Variation of the spectral index along the radio lobes at low frequency $\rm \alpha_{144MHz}^{614MHz}$ (top panel) and at high frequency $\rm \alpha_{1400MHz}^{4850MHz}$ (bottom panel). In both cases the spectral index shows a systematic and quick steepening from the inner region of the lobes towards the edges. Values have been extracted from the black rectangular regions shown in the middle panel of Fig. \ref{fig:spec_map}. The grey line represents a spectral index equal to $\rm \alpha=1$ as a reference. The 0 value on the x-axis corresponds to the innermost point in each lobe, where the flattest spectral index is measured. We note that one beam corresponds to five pixels. An average error bar is shown in the bottom left corner of both plots.}
\label{fig:spec_map_distr}
\end{figure}

To further quantify the spectral index variation throughout the lobes, we plotted its value along two slices drawn from the western and eastern lobe for comparison to \cite{roettiger1994} (Fig. \ref{fig:spec_map_distr}). The slices were chosen to extend from the flattest region within the lobes to the outermost edges (see Fig. \ref{fig:spec_map}, middle panel).     
Finally, using the spectral index maps described above, we computed a spectral curvature (SPC) map defined as $\alpha_{\rm 1400MHz}^{\rm 4850MHz}$-$\alpha_{\rm 144MHz}^{\rm 614MHz}$ (Fig. \ref{fig:spec_map}, right panel).

\subsubsection{Spectral age maps}
\label{pixelage}

The curvature of the radio spectrum of a source is dictated by the amount of radiative losses that the particles have experienced, and therefore it is directly related to the plasma age following the equation

\begin{equation}
t_s=1590\frac{B^{\rm 0.5}}{(B^2+B_{\rm CMB}^2)\sqrt{\nu_{\rm b}(1+z)}} \\ ,
\label{eqtime}
\end{equation}

\noindent where $t_s$ is the age in Myr, $B$ and $B_{CMB}=3.25\cdot(1+z)^2$ are the magnetic field and inverse-Compton equivalent field in $\rm \mu G$, $\nu_{\rm b}$ is the break frequency in GHz, and $z$ is the redshift. From this expression it is clear that the age $t_s$ directly scales with the magnetic field strength $B$.

In order to estimate the age of a source, two approaches can be taken. The historical approach is based on the spectral break measurement and on the use of the analytical equation shown above. The second approach instead consists of fitting the observed radio spectrum with a modelled spectrum that is obtained by numerical integration of the equations that describe the radiative losses of the plasma through synchrotron emission and inverse-Compton scattering with the CMB.
We here used \texttt{BRATS}, which follows this second approach (for a full derivation of the underlying equations, we refer to \citealp{harwood2013}).

In particular, to describe the spectral shape of a source, various models have been proposed that rely on different initial assumptions. One category of models assumes that the electrons are accelerated in a single event at a time $\rm t_0$ with an energy distribution equal to $N(E, t) = N_0 E^p$ (where $p$ is the particle energy power index), which translates into a power-law spectrum of the form $S \propto \nu^{-\alpha_{inj}}$ (where ${\alpha_{inj}}$ is the injection spectral index and has typical values in the range 0.5-0.8). As the particles age, the high-frequency tail of the radio spectrum undergoes a steepening as a result of preferential cooling of high-energy particles. 

The Kardashev-Pacholczyk model (KP, \citealp{kardashev1962}; \citealp{pacholczyk1970}) and the Jaffe-Perola model (JP, \citealp{jaffe1973}) are two classical models of this kind and assume a uniform magnetic field distribution in the source. The main difference between the two concerns the micro-physics of the electron population. While in the KP model the pitch angle (the angle between the velocity vector and the magnetic field) of individual electrons is considered to be constant, the JP model assumes a more realistic situation where single particles are subject to many scattering events that randomise their pitch angle. In practice, this is equivalent to assuming a timescale for the isotropisation of the electrons that is much longer than the radiative timescale. This different assumption naturally leads to differences in the curvature of the spectrum. In particular, for a given age, the KP model is relatively flatter that its JP counterpart at high frequencies. The reason are high-energy electrons at small pitch angles, which are able to radiate at higher frequencies.

A third model is the Tribble model \citep{tribble1991, tribble1993}, which includes a more realistic magnetic field distribution. In particular, it assumes the magnetic field to be spatially non-uniform, which in the weak-field strong-diffusion case (i.e. free streaming) can be described by a Maxwell-Boltzmann distribution within each volume element of the lobe. This has been expanded to an implementable form by \cite{hardcastle2013b} and \cite{harwood2013}.

When this approximation does not hold, there is a second category of models that assume a continuous injection of particles throughout the lifetime of the source. These are constructed by summing individual JP or KP spectra related to particle populations of different ages. In particular, the continuous injection (CI) model \citep{pacholczyk1970} best describes active sources in which the injection of fresh particles is still ongoing, while the so-called CIOFF or KGJP/KGKP model \citep{komissarov1994} assumes that the particle injection in the source is continuous for a certain amount of time and then ceases.

The assumption of a single injection made in the JP, KP, and Tribble models can work reasonably well for resolved spectral studies because on small scales, particles can most likely be considered as being part of the same acceleration event. In particular, in the following analysis we performed a spatially resolved (pixel-by-pixel) spectral modelling considering the JP and the Tribble models because they implement a more realistic physics. As a first step, we derived the best value for the injection index, $\rm \alpha_{inj}$ to be used in the final age estimate. To do this, we performed a series of fitting iterations over the entire source using the JP and the Tribble models. While keeping all the other parameters fixed, we firstly varied $\rm \alpha_{inj}$ over a grid ranging between 0.5 and 1 with a step size of 0.05. Secondly, we refined our grid to a step size of 0.01 around the previous minimum. The best-fit value over the entire source obtained using the two models is equal to $\rm \alpha_{inj}$=0.57, consistent with the low-frequency spectral index measured in the western hotspot.

\begin{figure}[htp!]
\center
\includegraphics[width=0.35\textwidth]{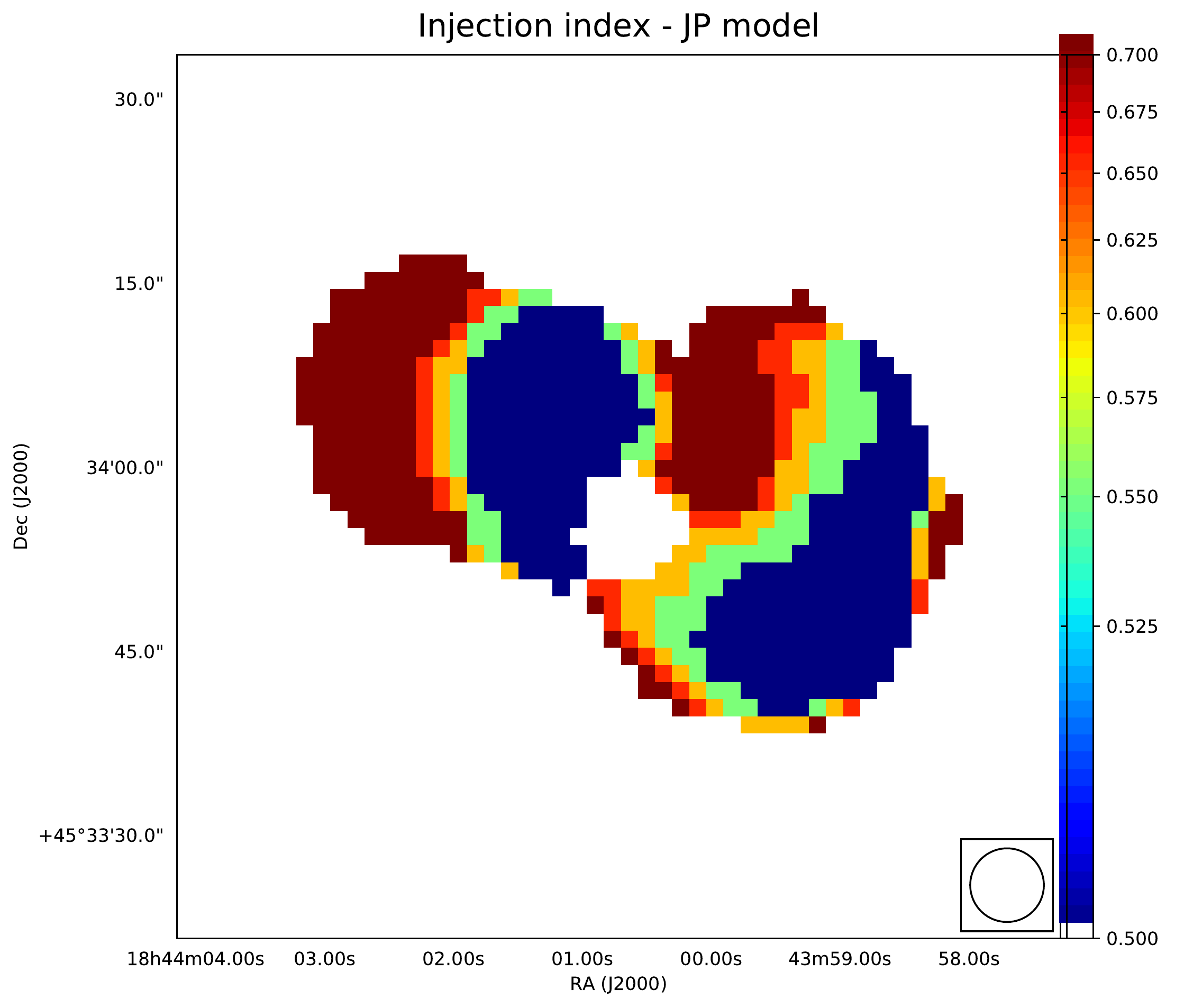}
\caption{Pixel-by-pixel best-fit injection index value obtained using the JP model. The value of the injection index appears to be correlated with the position within the radio lobes of the radio galaxy. }
\label{fig:inj}
\end{figure}

We note that for both models, the best value of $\rm \alpha_{inj}$ is strongly correlated with the position in the radio lobe, with values up to $\rm \alpha_{inj}\sim$ 0.7 in the outer lobes and values of $\rm \alpha_{inj}\sim$ 0.5 in the inner regions of the lobes (see Fig. \ref{fig:inj}). The observed systematic spatial trend appears to suggest an intrinsic variation of the plasma properties in the different regions of the source, which might be related to the two claimed jet episodes. However, given the large error bars on the low-frequency spectral index, it remains challenging to confirm this trend.

\begin{table*}[htp!]

\small
\caption{Model fitting results. Col. 1 lists the model we used for the fitting, Col. 2  the magnetic field in $\rm \mu$G, Col. 3 the injection index, Col. 4 the mean $\rm \chi^2_{Red}$ over all regions, Col. 5 the pixel with maximum age in Myr, Col. 6 the pixel with minimum age in Myr, Col. 7 the number of pixels whose $\rm \chi^2$ values fall within a specific confidence range, Col. 8  shows
whether the goodness-of-fit over all the regions is rejected, and Col. 9 lists the confidence level at which the model can be rejected over the entire source.}
\onehalfspacing
        \centering
                \begin{tabular}{c c c c c c c c c c c c c}
                \hline
                \hline
                Model & B & $\rm \alpha_{inj}$  & Mean $\chi^2_{Red}$ & Max age & Min age & \multicolumn{5}{c}{Confidence bins} & Rejected & Median\\
        
                & [$\rm \mu$G] & & & [Myr] & [Myr] &<68 & 68-90 & 90-95 & 95-99 & $\rm \geq$99 & & confidence\\
                \hline
                JP & 15.8 & 0.57 &  2.13 & $\rm 16.78^{+2.68}_{-2.14}$& $\rm 1.99^{+0.09}_{-0.62}$ & 246 & 145 & 75 & 138 & 275 & No & <68\\
                Tribble & 15.8 & 0.57 & 2.08 & $\rm 18.48^{+3.07}_{-1.92}$ & $\rm 1.99^{+0.44}_{0.58}$ & 249 & 150 & 65 & 129 & 286 & No & <68\\
                \hline
                JP & 3 & 0.57 &  2.11 & $\rm 82.05^{+13.79}_{-9.44}$ & $\rm 9.05^{+2.65}_{-2.29}$ & 249 & 148 & 71 & 137 & 274 & No & <68\\
                Tribble & 3 & 0.57 &  2.06 &$\rm 90.95^{+0.0}_{-11.39}$ & $\rm 9.95^{+2.09}_{-3.04}$&  255 & 157 & 64 & 125 & 278 & No & <68\\
                
                \hline
                
            \hline                  
                \hline                  
                \end{tabular}          
        \label{tab:fitres}          
\end{table*}

\begin{figure*}[!htp]
\minipage{0.33\textwidth}
  \includegraphics[width=1\textwidth]{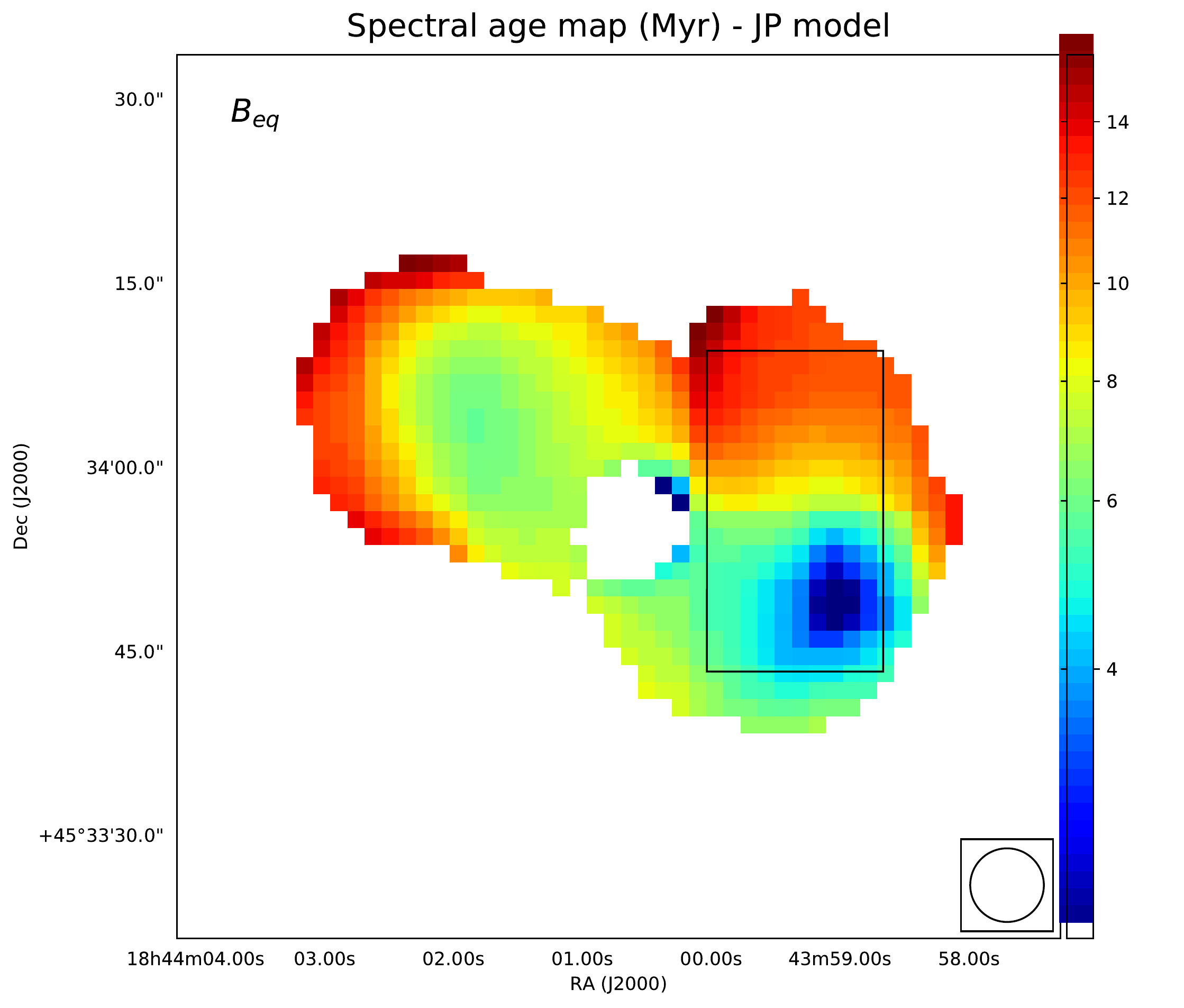}
\endminipage\hfill
\minipage{0.33\textwidth}
  \includegraphics[width=1\textwidth]{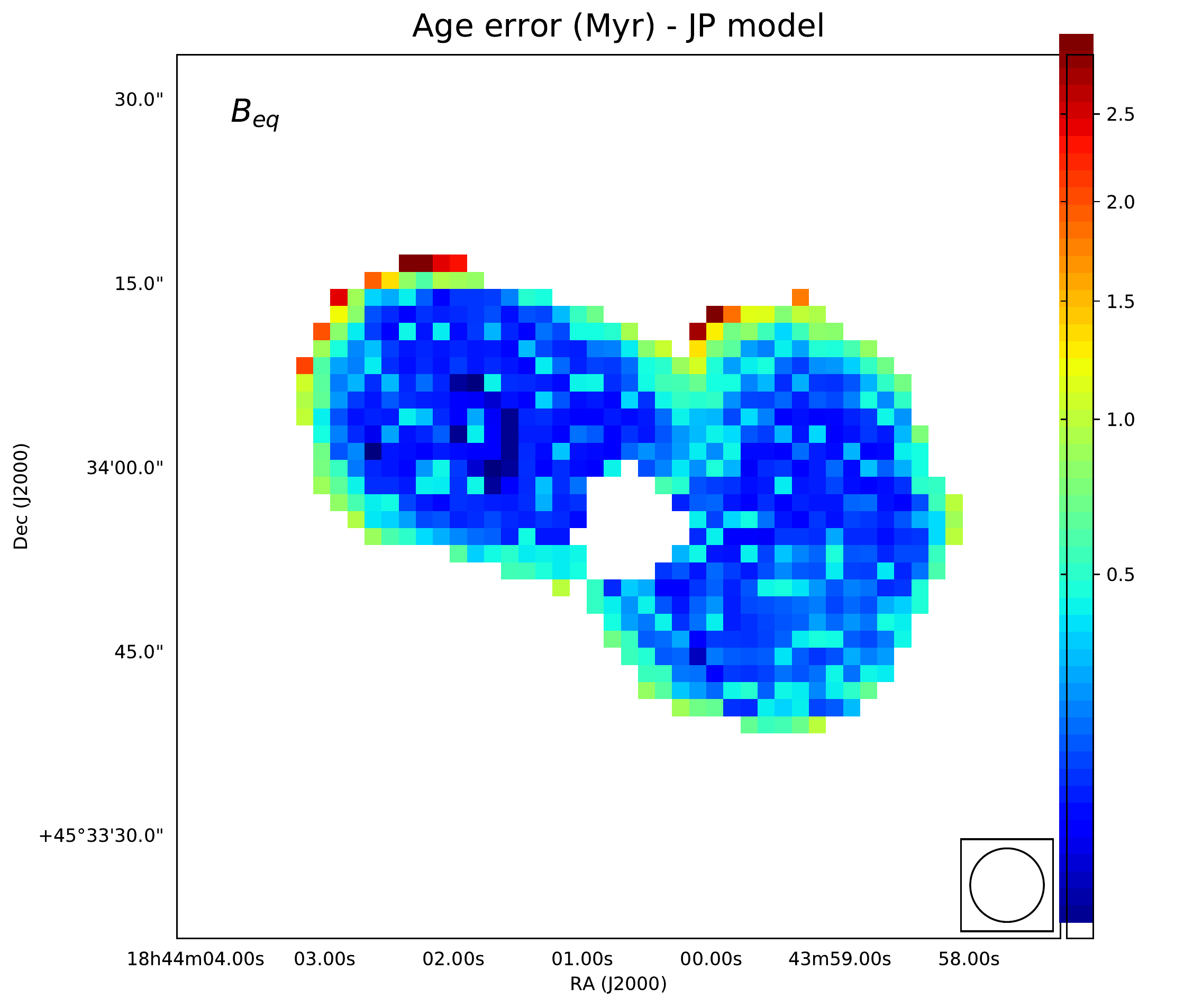}
\endminipage\hfill
\minipage{0.33\textwidth}
  \includegraphics[width=1\textwidth]{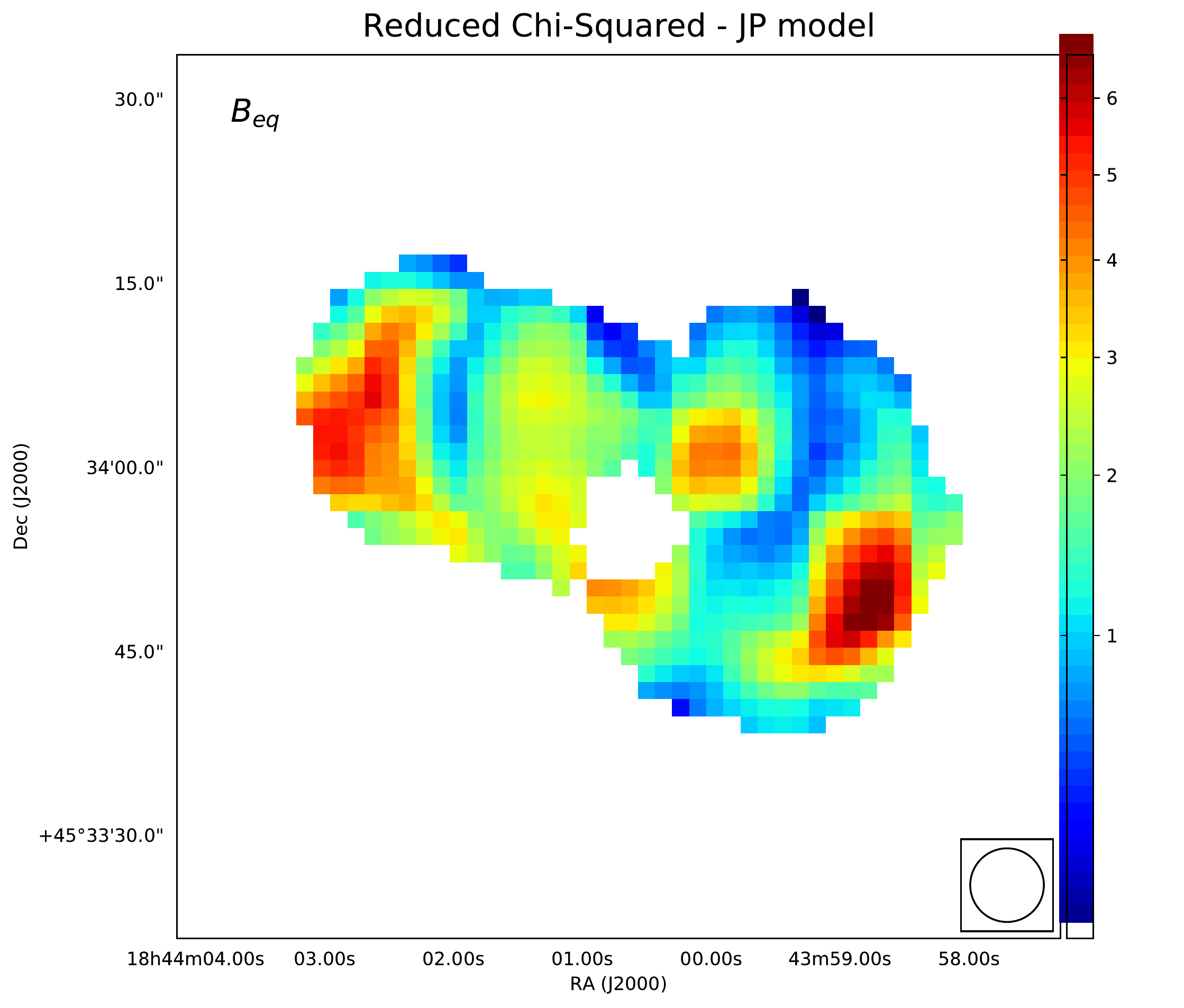}
 \endminipage\hfill 
  \minipage{0.33\textwidth}
  \includegraphics[width=1\textwidth]{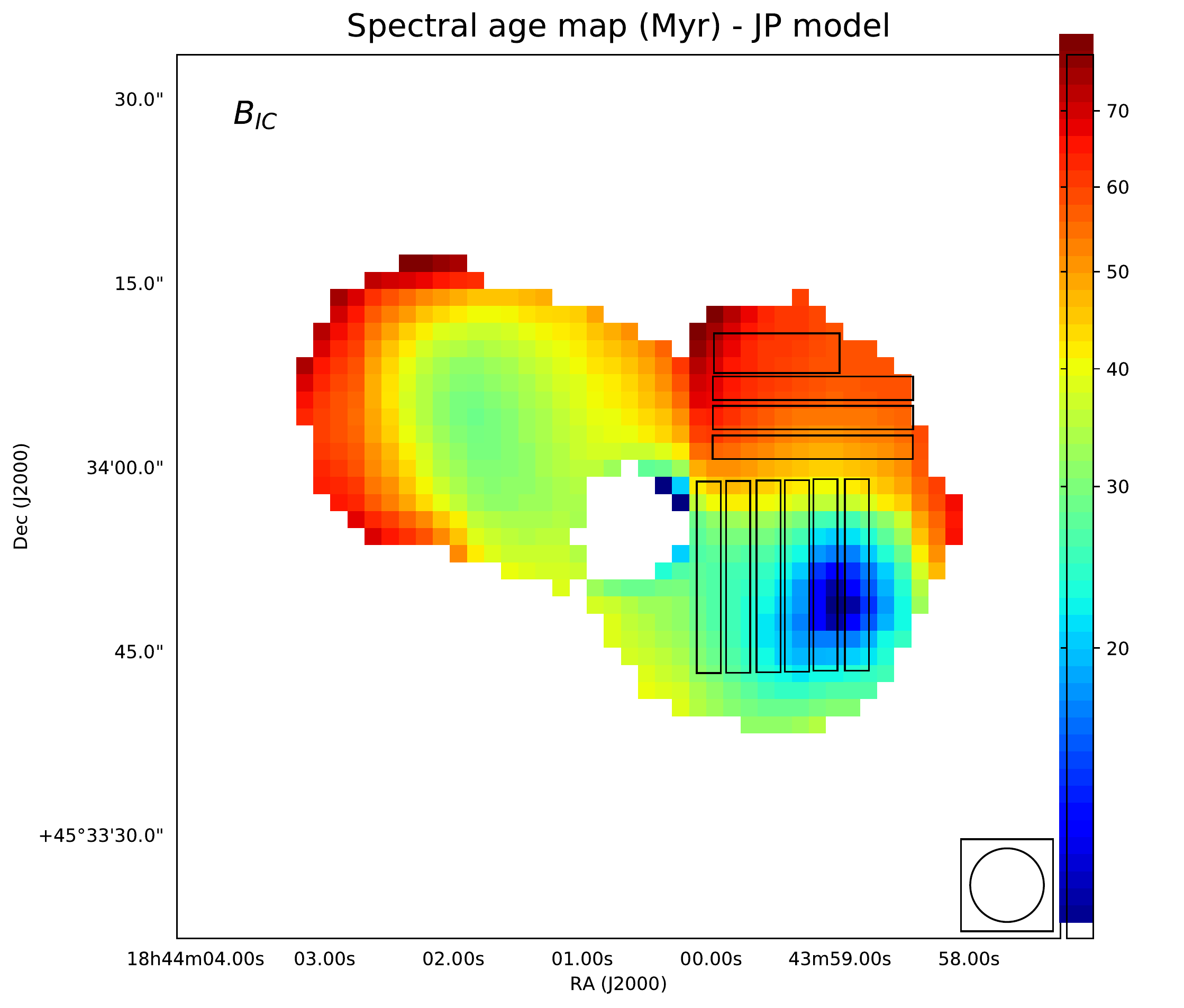}
\endminipage\hfill
\minipage{0.33\textwidth}
  \includegraphics[width=1\textwidth]{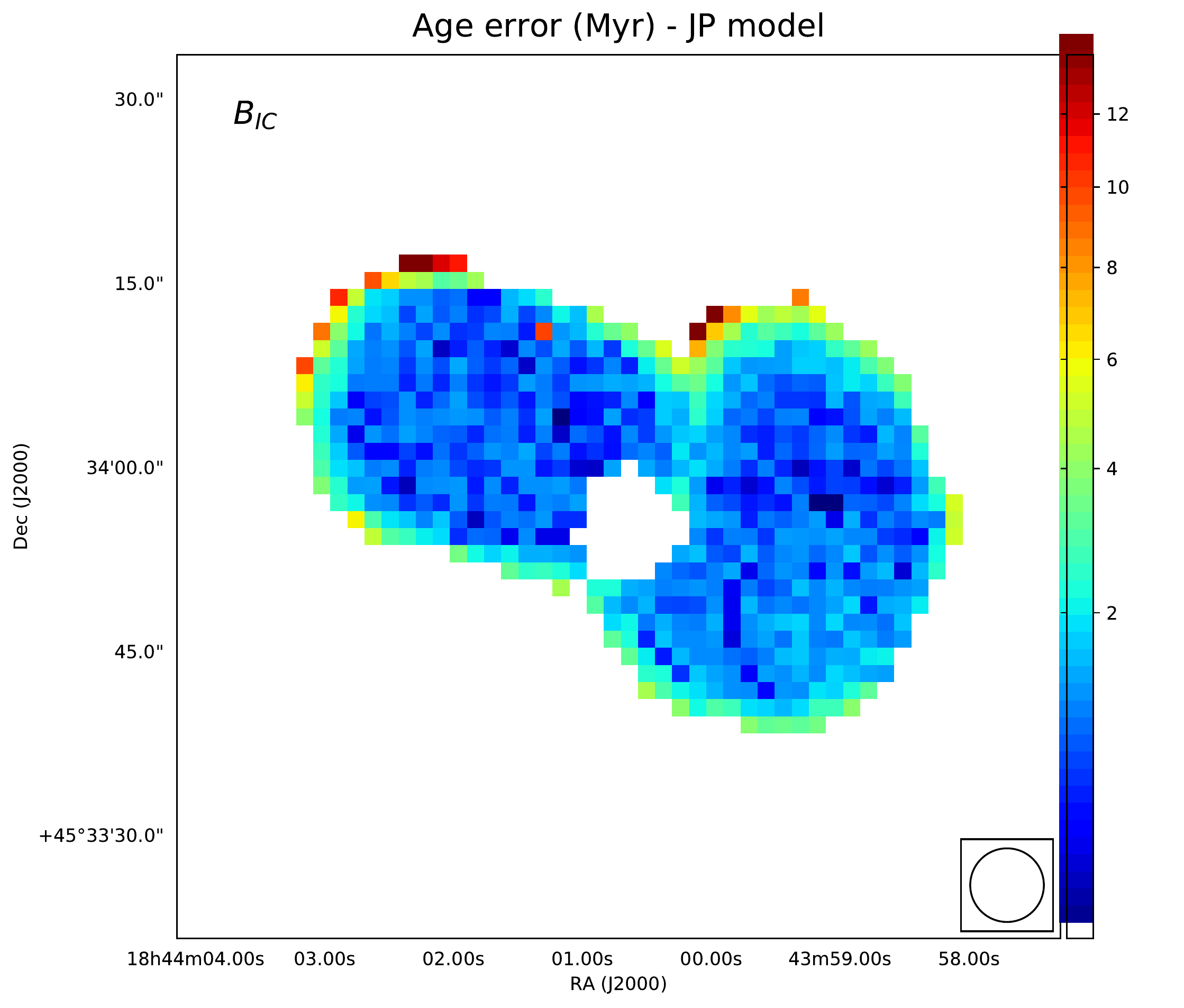}
\endminipage\hfill
\minipage{0.33\textwidth}
  \includegraphics[width=1\textwidth]{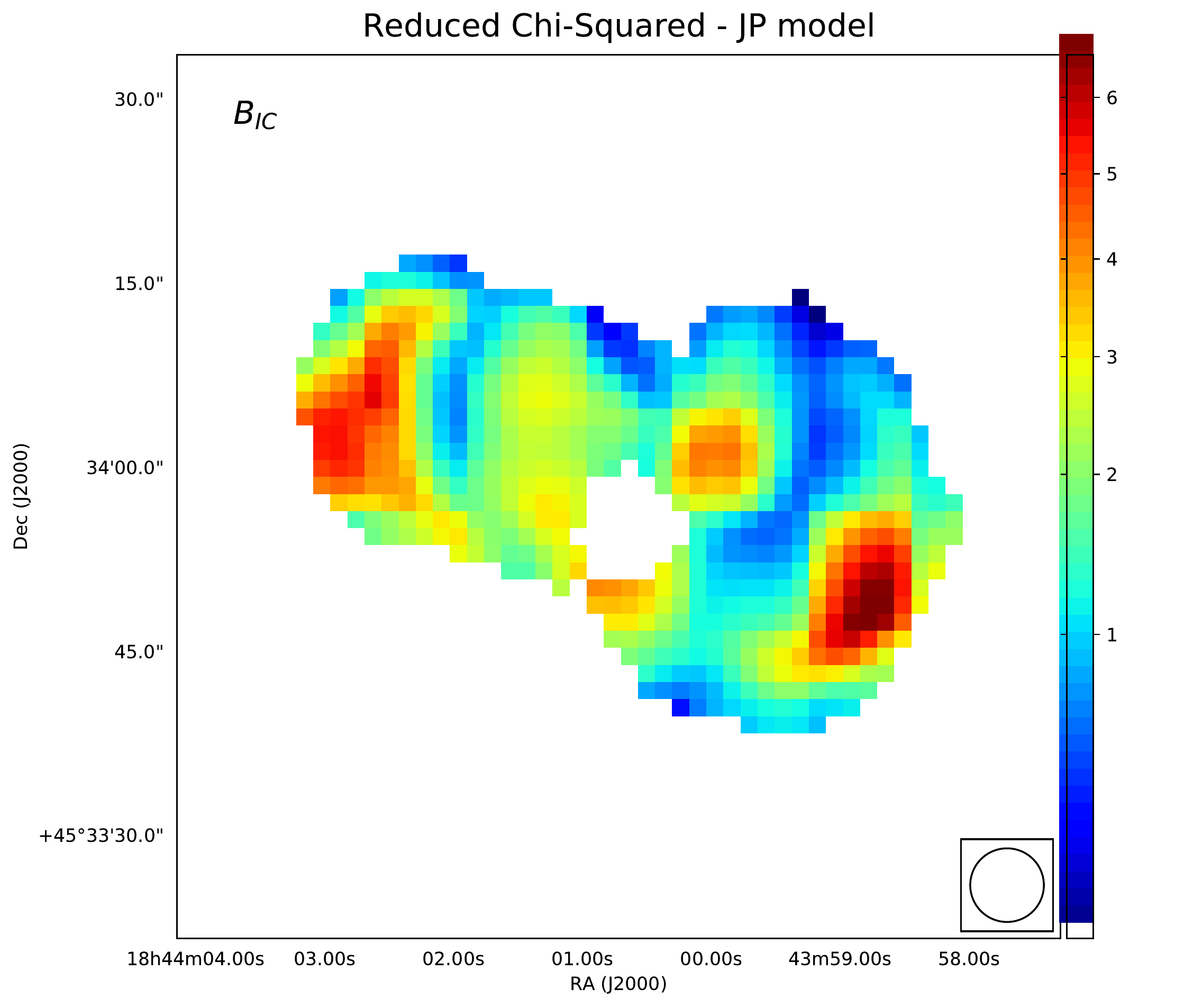}
  \endminipage\hfill
  \caption{Spectral age maps of the source 3C388 (left) and relative age error maps (middle) and reduced $\rm \chi^2$ maps (right) obtained using the JP model. In the top row, maps have been produced assuming a magnetic field equal to the equipartition magnetic field $\rm B_{eq}$ and in the bottom row equal to $\rm B_{IC}$. A clear increase of spectral age is apparent from the inner lobes towards the source edges. The reduced $\rm \chi^2$ maps also show that the goodness-of-fit is not uniform across the source, which might be caused by mixing of different particle populations. The black rectangle in the top left panel represents the region from which pixel-based values have been used for the colour-colour diagram analysis presented in Sect. \ref{color} (circles in the plot). The black rectangles in the bottom left panel represent the regions from which integrated values have been computed for the colour-colour diagram analysis presented in Sect. \ref{color} (squares and diamonds in the plot). }
\label{fig:age_map}
\end{figure*}

\begin{figure}[htp!]
\centering
  \includegraphics[width=7cm]{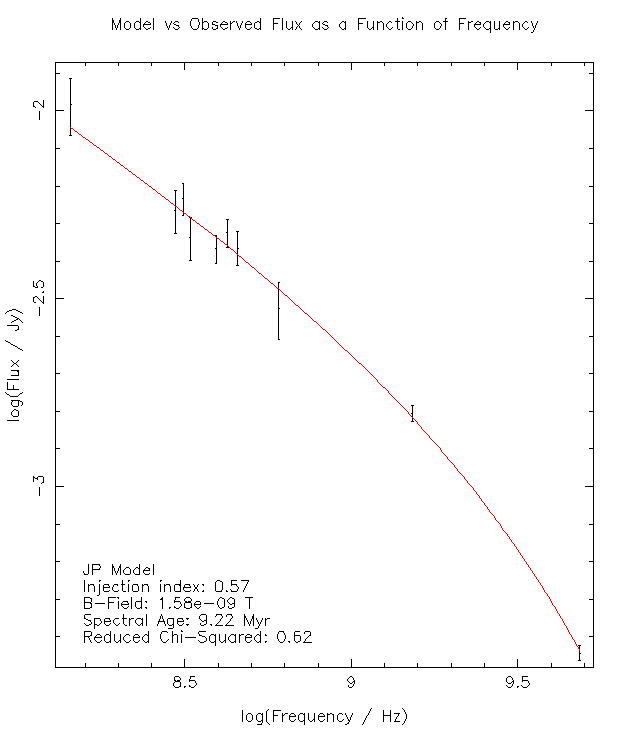}

  \caption{Representative radio spectrum (flux density vs. frequency) for an individual pixel with a good fitting result. Black points are observational data, and the red solid line is the best fit using the JP model. Model parameters, ages (in Myr), and statistics are shown in the bottom left corner of the panel. }
\label{fig:spec-pix}
\end{figure}

We therefore decided to use as injection index the best-fit value over the entire source equal to $\rm \alpha_{inj}$=0.57 for the final spectral age derivation. We ran the final model fitting iteration using the JP and Tribble models and the two magnetic field values described in Sect. \ref{magn} ($\rm B_{eq}=15.8 \ \mu G$ and $\rm B_{IC}=3 \ \mu G$). The final fitting results are presented in Table \ref{tab:fitres}. In Fig. \ref{fig:age_map} we also show the final spectral age maps obtained using the JP model with respective error maps and $\rm \chi^2_{red}$ (8 degrees of freedom). One representative spectral plot (flux density versus frequency) for an individual pixel with good fitting results is also shown in Fig. \ref{fig:spec-pix} for illustration purposes. The fits obtained using the Tribble model provide age and $\rm \chi^2_{red}$ distributions that are comparable with the JP model within the errors, and therefore we do not show their maps here. 

 Table \ref{tab:fitres} shows that the overall fitting results are poor with a mean $\rm \chi^2_{red} \sim$ 2 over the entire source. None of the models can be rejected at the 68\%\ confidence level over the entire source, and we note that a significant number of regions can be rejected at the 95\% or 99\%\ confidence level. These regions are also clearly visible as red pixels in the $\rm \chi^2_{red}$ maps in Fig. \ref{fig:age_map} 
This suggests that the spectral shape is not constant
throughout the source, as is further discussed in Sect. \ref{dissage}.

Finally, following the injection index best-fit distribution within the lobes (see Fig. \ref{fig:inj}), we investigated how these results would vary if we assumed an injection index equal to 0.7 for the outer lobe regions and an injection index equal to 0.5 for the inner lobe regions. With these new sets of input parameters, we find deviations in the final ages up to a few Myr. These values lie well within the considered final error on the ages. For this reason, we do not report these results here and in the following analysis only consider the ages obtained using a common injection index throughout the source equal to $\rm \alpha_{inj}$=0.57, as described above.

\subsubsection{Colour-colour diagram}
\label{color}

To further investigate the spectral properties in different regions of the source we used the colour-colour diagram ($\rm \alpha_{high}$ versus $\rm \alpha_{low}$, \citealp{katzstone1993, katzstone1997, vanweeren2012, shulevski2015}). This plot is useful to inspect and compare the curvature of the radio spectrum in different regions or pixels of the source, independently of the assumption on magnetic field and on the presence of adiabatic compression or expansion. These mechanisms, indeed, can only cause a shift of the spectrum in frequency but do not affect its actual shape.
Therefore, it allows us to distinguish among different radiative models, as well as to probe the presence of multiple particle populations.  

\begin{figure}[htp!]
\centering
  \includegraphics[width=9.5cm]{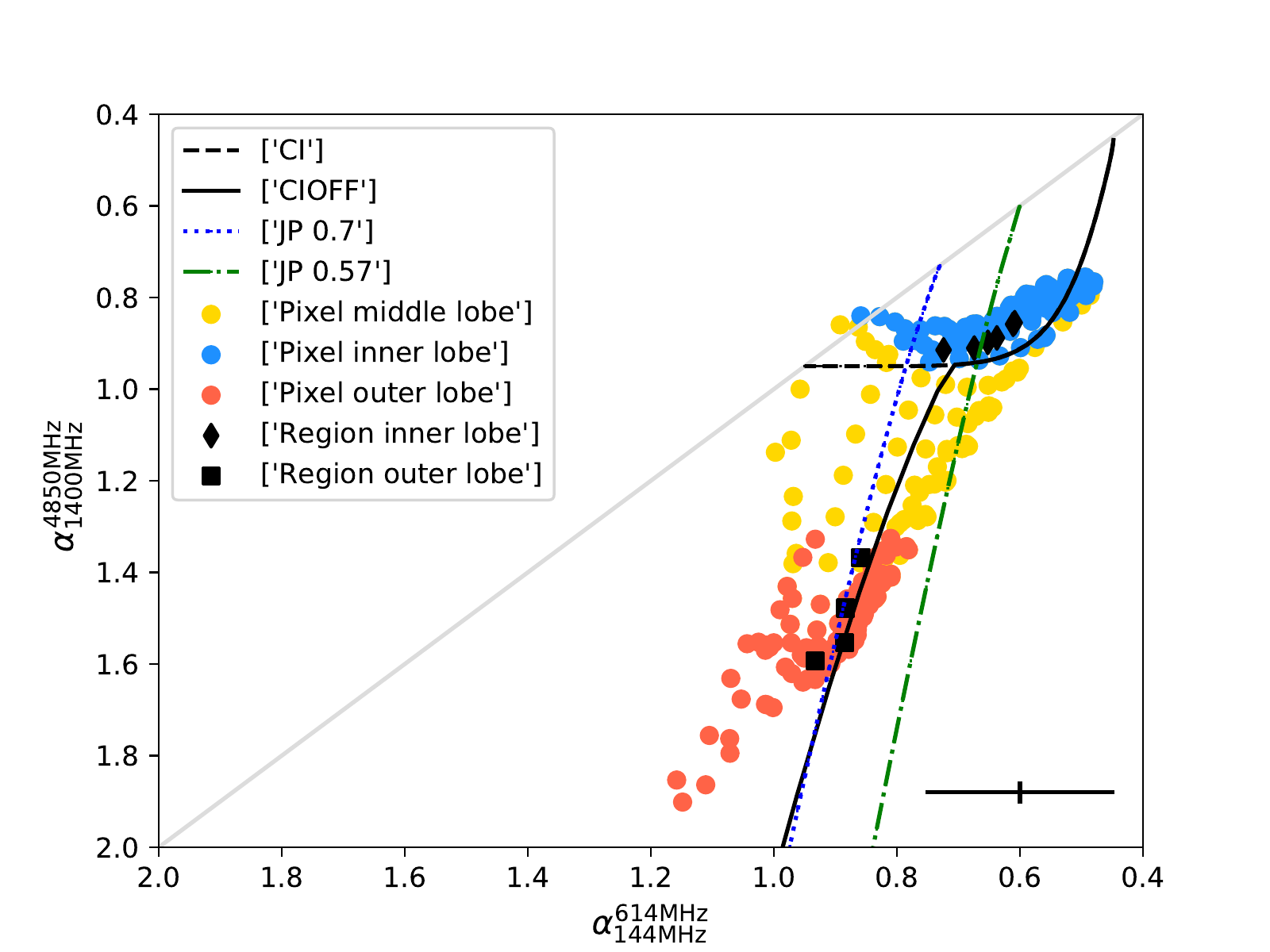}

  \caption{Colour-colour diagram of the western lobe showing that the spectral shape of the plasma located in different regions follows different radiative model curves. Pixel-based values are shown as circles (blue shows the inner lobe, red the outer lobe, and yellow the middle region of the lobe). Black diamonds and squares are the values computed by integrating the flux density in the regions shown in the bottom left panel of Fig. \ref{fig:age_map} for the inner and outer lobe, respectively. The grey solid line represents $\alpha_{\rm 144MHz}^{\rm 614MHz}$=$\alpha_{\rm 1400MHz}^{\rm 4850MHz}$. The other lines indicate the spectral index evolution of ageing particles for various models. In particular, the black dashed line represents a CI model with $\rm \alpha_{inj}=0.45$, and the black solid line represents a CIOFF model with $\rm \alpha_{inj}=0.45$ B=3 $\rm \mu G$ and $\rm t_{on}=20 \ Myr$. The blue dotted line represents a JP model with $\rm \alpha_{inj}=0.7,$ and the green dash-dotted line represents a JP model with $\rm \alpha_{inj}=0.57$. The average error bar on the spectral indices is shown as a reference in the bottom right corner.}
\label{fig:color}
\end{figure}

In Fig. \ref{fig:color} we show the colour-colour diagram $\alpha_{\rm 1400MHz}^{\rm 4850MHz}$ vs $\alpha_{\rm 144MHz}^{\rm 614MHz}$ obtained for the western lobe. 
We restrict the analysis to this lobe because its larger extension allows for a more detailed analysis of the difference between the inner and outer lobe. As circles we show the pixel-based values extracted from the region shown in Fig. \ref{fig:age_map} top left panel, coloured according to their position within the lobe. 
Black diamonds and squares represent instead the values computed by integrating the flux density in the regions shown in Fig. \ref{fig:age_map}, bottom left panel, for the inner and outer lobe, respectively. 
Finally, we plot the lines corresponding to some spectral models as a reference: 1) a JP model with $\rm \alpha_{inj}$=0.57 (dash-dotted green line), 2) a JP model with $\rm \alpha_{inj}$=0.7 (dotted blue line), 3) a CI model with $\rm \alpha_{inj}$=0.45 (dashed black line), and 4) a CIOFF model with $\rm \alpha_{inj}$=0.45 (solid black line). For this last model we fixed the magnetic field to B=3 $\mu G$ and the active phase of the source to 20 Myr. The plot clearly shows that the spectral shape is not uniform throughout the source. A full discussion of the observed trends is presented in Sect. \ref{dissage}.

\section{Discussion}
\label{discussion}

\subsection{Spectral index distribution}
\label{dissindex}

The spatial distribution of the spectral index that we computed in the range 1400-4850 MHz (see Fig. \ref{fig:spec_map}, middle panel) is consistent with previous results by \cite{roettiger1994}. By studying the spectral index variation in the radio lobes at high frequency (see Fig. \ref{fig:spec_map_distr}, bottom panel), we also confirm the rapid steepening towards the edges that was identified previously. The gradient we observe is smoothed with respect to what was presented by \cite{roettiger1994} by the effect of a larger beam in our images (a factor 4 larger).
In particular, in the western lobe the spectral index varies from $\rm \alpha_{1400MHz}^{4850MHz}$=0.75 in the centre of the compact hotspot-like emission to $\rm \alpha_{1400MHz}^{4850MHz}$=1.56 in the most external edge of the radio lobe. In the eastern lobe, the spectral index varies from $\rm \alpha_{1400MHz}^{4850MHz}$=0.89 in the centre of the diffuse hotspot to $\rm \alpha_{1400MHz}^{4850MHz}$=1.51 in the most external edge of the radio lobe.

The spectral index distribution at low frequencies is presented in this work for the first time (Fig. \ref{fig:spec_map}, left panel) and shows similar trends to the one observed at high frequency, with flatter spectral indices in the vicinity of the centre of the radio lobes and steeper spectral indices towards the edges of the lobes. 
However, as expected from spectral evolution models, the spectral indices at low frequency are systematically flatter than those at high frequency over the entire source. 
Figure \ref{fig:spec_map_distr} (top panel) shows that the variation of the spectral index $\rm \alpha_{144MHz}^{614MHz}$ across the two lobes is milder than what is observed at higher frequencies, with values ranging from $\rm \alpha_{144MHz}^{614MHz}$ $\sim$0.55 to $\sim$0.9 in the western lobe and from $\rm \alpha_{144MHz}^{614MHz}$ $\sim$0.60 to $\sim$1.14 in the eastern lobe.

Based on the new lo- frequency data, we were also able to investigate the spectral curvature distribution across the lobes (see Fig. \ref{fig:spec_map}, right panel), from which the overall spectral shape in different regions of the source can be better appreciated. In the inner regions of both lobes, the spectral curvature has values in the range $\rm 0.1\leq SPC \leq 0.3$, which are compatible with the plasma being still accelerated by the jets. At the outer edges of both lobes, and especially in the western lobe, the spectral curvature instead increases significantly, with values up to SPC=0.7-0.8. This extreme curvature is typical of old ageing plasma that is not replenished with new freshly injected particles, supporting the idea of a remnant electron population. We also note that in some pixels at the lobe edges, the spectral curvature is highly reduced. Whether this effect is real is difficult to assess because the image fidelity at the source border is often poor (see \citealp{harwood2013, harwood2015}).

In light of the results at low and high frequency, the spectral behaviour of 3C388 is difficult to reconcile with what was observed in classical radio galaxies. FR~II sources typically show the flattest spectral indices at the lobe edges in correspondence of the hotspots, where the particle acceleration occurs, and the steepest spectral indices in the regions surrounding the core, where the oldest accelerated particles are located (e.g. \citealp{carilli1991, orru2010, mckean2016}). In FR~I radio galaxies the spectral index may both steepen from the core outwards or from the lobe outer edges inwards, depending on the source morphology \citep{parma1999}. Lobed FR~I radio galaxies tend to show the flattest spectral index values all along the jets up to the lobe edges and the steepest spectral index values in the surrounding regions \citep{laing2011}. Instead, plumed (or tailed) FR~I radio galaxies have the flattest spectral indices in the core regions, which then become increasingly steeper with distance \citep{laing2008, heesen2015}. 

None of the spectral classes described above completely matches what we observe in 3C388, which remains a special case. The only source in the literature showing a similar spectral trend to 3C388 is Hercules A (3C348), where a sharp gradient in spectral index is also observed between the inner brighter regions of the lobes and the surrounding diffuse ones \citep{gizani2003}. In the case of 3C348, high-resolution radio images clearly show that this trend can be attributed to the presence of a new-born jet pushing away and compressing the old lobe material from a previous outburst. A similar scenario can therefore be suggested for 3C388. 

As an alternative, the observed spectral distribution might be dictated by a combination of source bending and projection effects. While this scenario would require a very peculiar geometry, as discussed in \cite{burns1982}, this possibility cannot be completely discarded. Recent studies also suggest that atypical morphologies and spectral distributions in radio galaxies are likely attributable to these effects \citep{harwood2019}.

\subsection{Spectral ages and models}
\label{dissage}

As expected, the spectral age distribution follows the observed spectral index distribution, with younger ages in the inner lobes and older ages at the lobe edges (see Fig. \ref{fig:age_map}). Again we underline that in contrast to typical FR~IIs where the youngest regions correspond to the hotspots at the very edge of the lobes, in 3C388 the youngest ages are observed in the location of the compact hotspot in the western lobe and diffuse hotspot in the eastern lobe, as defined by \citealp{roettiger1994}), both embedded in the lobes.

Table \ref{tab:fitres} shows that the results obtained using the two different radiative models (Tribble and JP) are consistent within the errors. 
Because of this, we only refer to the JP model results in the discussion below. We stress that the largest uncertainties on the age estimate are dictated, as expected, by the different assumptions on the magnetic field. The absolute age values increase by about a factor of 5 when we use $\rm B_{IC}=3 \ \mu G$ ($\rm t_{max}\sim 80$ Myr - $\rm t_{min}\sim9$ Myr) with respect to the simple equipartition assumption $\rm B_{eq}=15.8 \ \mu G$ ($\rm t_{max}\sim 16$ Myr - $\rm t_{min}\sim 2$ Myr).

Because the oldest age found in the resolved spectral analysis can be considered representative of the first particle acceleration in the source, we can infer that the total age of the source is $\lesssim$80 (for the most realistic assumption of magnetic field close to the inverse-Compton limit). This value is compatible with a dynamical age of the radio source equal to $<$65 Myr, estimated to first order by \cite{kraft2006} by assuming the lobes to be buoyant bubbles expanding in the ambient medium at a velocity equal to half the sound speed.

Another interesting point to highlight is that as described in Sect. \ref{spec}, the overall fitting results over the entire source cannot be rejected with only 68\%\ confidence, meaning that there is a significant number of regions that shows poor results. This can be appreciated from Table \ref{tab:fitres} and the right panels of Fig. \ref{fig:age_map}, which clearly show that there are regions with $\rm \chi^2_{Red}$ values up to 10. The location of these poorly fitted regions does not correlate with either the outer or inner lobes. It mainly corresponds instead to the hotspot in the western lobe and the outer edges of the eastern lobes.

This finding appears to suggest that in some regions in the lobes of this radio galaxy, the physical conditions of the plasma cannot be described by the spectral models that we used. A possible cause of the poor fitting results may be strong mixing of different particle populations within the lobes of the source. This particle mixing would invalidate the simple assumption of a single injection event, as used in the JP and Tribble models. 

This possibility has previously been discussed from an empirical stand point (e.g. \citealp{harwood2017, harwood2019, mahatma2019}) and is also supported through modelling considerations and numerical simulations \citep{rudnick2002, turner2017}, which show that the mixing of different aged electrons strongly affects the spectrum at each point of the radio source leading to poor spectral age estimates. This mixing may have an observational origin, when the resolution of the observation is not high enough to distinguish different particle populations or in case of projection effects (see \citealp{harwood2019}), and also an intrinsic origin, when there is an actual mixing of particles that are accelerated by different events (i.e. having different physical properties such as injection index and magnetic field) or at different times.

In FR~II radio galaxies, for example, this may be particularly relevant in the backflow region, where freshly injected electrons from the hotspots are carried back towards the AGN core, causing a significant mixing of particle populations. Mixing may become even more relevant if we think that in 3C388 newly started jets are expanding in old ageing lobes, as suggested by the restarting AGN scenario \citep{roettiger1994}. All this reflects on the uncertainties of the derived spectral age, which should therefore be taken with care.

In this context, the colour-colour diagram gives us some further insights into the spectral shape of the plasma in different regions of the source 3C388.  
From the plot shown in Fig. \ref{fig:color}, it is clear that despite the large error bars, the points related to the middle or outer and inner lobe follow two very distinct trends. This conclusion remains valid when the pixel-based (circles) and the region-based information (squares and diamonds) is used and again supports the idea that we study two distinct electron populations that experienced very different amounts of radiation losses.

In particular, the points related to the inner lobe quite impressively follow the CI model trend. 
We note, however, that the points do not reach the single power-law (grey) line in correspondence of the injection value, possibly because of particle mixing. In contrast, they touch the single power-law (grey) line at the bottom of the curve, showing an upturn with respect to the plotted CI model. The fact that the CI model best describes the spectral shape in the inner lobe explains why poor fitting results using the JP model are obtained in correspondence of the hotspot (see Fig. \ref{fig:age_map} right).

The points related to the middle and outer lobe instead clearly deviate from the CI curve, showing much stronger steepening. The pixel-based points show a large scatter, which is hard to reconcile with one single model. This might be an indication of different physical conditions in different regions of the plasma, but it might also simply be the effect of the large error bars. 
However, the region-based points show an impressively good agreement with a CIOFF model with an active time of $\rm t_{on}$=20 Myr with B=3 $\rm \mu G$. This may be suggesting that the remnant plasma of the outer lobe was produced by an episode of jet activity similar to the currently ongoing episode (which matches the CI trend described above well) and that switched off after 20 Myr. A more thorough investigation of the jet activity timescales is presented in the next section.

\subsection{Duty cycle}
\label{duty}

In this section we investigate the duty cycle of 3C388 under the assumption that the discontinuity observed in the spectral distribution is not purely due to projection effects, but indicates a restarting jet activity, as proposed in the case of Hercules A (see discussion in Sect. \ref{dissindex}). As for the analysis of the colour-colour plot, we  focus on the western lobe, where the remnant plasma is much more extended and allows for a more detailed analysis. To derive the duration of the older jet activity phase, we used the same approach as in \cite{shulevski2017}: we computed the difference between the age of the oldest (82.05 Myr for $\rm B_{IC}$ and 16.78 Myr for $\rm B_{eq}$) and the youngest particle population (50.95 Myr for $\rm B_{IC}$ and 10.38 for $\rm B_{eq}$) measured in the outer western lobe (the remnant lobe). The duration of the first phase of jet activity found in this way is $\rm t_{1, on, B_{IC}}\sim$30 Myr and $\rm t_{1, on, B_{eq}}\sim$7 Myr, respectively.

We note that a reliable measure of the age of the youngest electron population in a remnant lobe comes from the region where the particle acceleration occurred during the active phase (e.g. a fading hotspot). In this region we can measure the age of the particles that were last accelerated before the switch-off. Unfortunately, in contrast to isolated remnant sources or double-double radio galaxies, where the outer lobes are well detached from the inner lobes, in 3C388 the particle acceleration region of the first period of jet activity is challenging to determine because it is likely currently mixed with the inner lobes. For this work we extracted the age of the youngest particles of the outer western lobe from a region including all the pixels showing a spectral curvature SPC>0.5, indicating very strong radiative losses typical of a remnant radio lobe. However, because of the abovementioned limitations, these numbers can only be considered as upper limits on the actual switch-off time.

To estimate the duration of the second episode of jet activity, we used the maximum age measured in the region closest to the core in the western lobe. We used this approach because, within the inner lobe, we observe the typical trend of FRII radio sources (see Fig. \ref{fig:age_map} left) in which the plasma closer to the nucleus is the oldest and becomes younger towards the hotspot where the current acceleration is taking place (e.g. \citealp{harwood2016}). In this way, we obtain an estimate of the second period of activity equal to $\rm t_{2, on, B_{IC}}\sim $30 Myr and $\rm t_{2, on, B_{eq}}\sim$6 Myr, respectively.

Using these values, we can compute a first-order estimate of the duty cycle of the radio jets in 3C388. 
In the case of $\rm B_{IC}$ , we find that the first jet episode lasted $\rm t_{1, on, B_{IC}}\gtrsim $30 Myr. 
This was followed by a period of inactivity that lasted $\lesssim $20 Myr, which is computed as the difference between the youngest age of the outer lobe, equal to 50.95 Myr, and the oldest age in the inner lobe, equal to 31.05 Myr. 
Finally, the current jet episode has lasted $\sim $30 Myr. When we instead use the values obtained assuming $\rm B_{eq}$ , the duty cycle is much shorter. Following the same procedure, we obtain that the first jet episode lasted $\gtrsim $7 Myr and was followed by an inactivity period of $\lesssim $4 Myr, followed by a second episode of jet activity $\sim $6 Myr long. 

In both cases the derived numbers provide a duty cycle of $\gtrsim$60\% defined as $\rm t_{on,1}/(t_{on,1}+t_{off}),$ in agreement with \cite{birzan2012}. The timescales of the jet activity derived here are summarised in Table \ref{tab:dutycycle}. We stress that because the most likely value of the magnetic field is close to the inverse-Compton limit (see Sect. \ref{magn-eq}), we consider the ages obtained with this values to be the closest to reality. Despite the variations obtained with different magnetic field assumptions and all the possible sources of error for the age described in Sect. \ref{dissage}, the duty cycle estimated for the source 3C388 seem to be consistent with the average values obtained from other studies of restarted radio galaxies.

\begin{table}[htp!]

\small
\caption{Duty cycle time-scales estimated using spectral age results obtained using the JP model on the Western lobe and different values of magnetic field (see Sect. \ref{dissage}).}
\onehalfspacing
	\centering
		\begin{tabular}{c c c }
		\hline
		\hline
		Jet phase & JP ($\rm B_{eq}$) & JP ($\rm B_{IC}$) \\
		\hline
		First episode $\rm t_{on,1}$ [Myr]& $\gtrsim $7& $\gtrsim$30\\	
		
		Inactive time $\rm t_{off}$ [Myr]& $\lesssim$4& $\lesssim$20 \\
		Second episode $\rm t_{on,2}$ [Myr]&>6&>30\\
		 Fractional duty cycle & $\gtrsim $60\% & $\gtrsim $60\%\\
		
	    \hline                  
		\hline	                
		\end{tabular}          
   	\label{tab:dutycycle}          
\end{table} 

In the specific case of Hercules A, the AGN has been claimed to have effectively ceased for a short period of $\sim$1 Myr and then restarted with a fluctuating jet activity of 250-800 kyr, which caused the rings that are observed in the radio morphology of the source \citep{gizani2003}. 

More in general, the typical estimated inactive period between two phases of activity varies in the range of a few Myr to a few tens of Myr for restarted radio galaxies of various morphologies, including the well-known double-double radio galaxies (e.g. \citealp{schoenmakers2000,saikia2006,konar2012, konar2013, shulevski2015, brienza2018}).  
Typically, the duration of this quiescent
phase is shorter or at most comparable to the duration of the previous phase of activity, which is usually found to be in the range few tens to few hundreds of Myr. The fact that we tend to detect sources with short quiescent periods is likely an observational bias that arises because the remnant plasma becomes quickly undetectable with current instruments even at MHz frequencies. This is becoming increasingly evident through recent observational campaigns of remnant radio galaxies detecting small fractions, up to 10\%, of these sources (see e.g. \citealp{brienza2017, godfrey2017, mahatma2018}), and also through radio galaxy modelling and simulations, which predict visibility timescales for the remnant plasma of the order of a few tens of Myr (e.g. \citealp{brienza2017, godfrey2017, hardcastle2018, english2019}).

Other constraints to the duration of the jet quiescent phase in radio galaxies come from the study of multiple generations of X-ray cavities at the centre of galaxy clusters. For a sample of 11 sources, \cite{vantyghem2014} found that the typical time interval between the two AGN outbursts that created the two pairs of X-ray cavities varies in the range $\sim$1-10 Myr, which is consistent with quiescent time estimates of restarted radio galaxies from radiative ages as discussed above. By comparing the outbursts intervals with the gas cooling time in the respective clusters, the authors find that the AGN in these systems restarted on a timescale of about a factor 3 shorter than the gas cooling time, making it an effective mechanism to suppress cooling flows.

Following a similar argument, \cite{kraft2006} showed that for 3C388 the jet mechanical power can easily quench the gas cooling if a duty cycle of only about 5\% with similar power is assumed. This requirement is much lower than the value we computed, which is equal to $\rm t_{on,1}/(t_{on,1}+t_{off})$=60\%. While it is impossible to predict whether the duty cycle that we probed will remain constant throughout the entire evolution history of the source, we can confirm that in this phase, the timescales of the jet activity in 3C388 are consistent with expectations from the X-ray analysis.

\section{Summary and prospects}

Because of its morphology and spectral index distribution at high frequency, the radio galaxy 3C388 has long been claimed to be a restarted radio galaxy. In this work, we have expanded the spectral study of the source to a much broader frequency range (144-4850 MHz) and estimated to first order the timescales of the jet activity. Here we summarise our main findings.
\medskip

(i) As expected by radiative evolution models, the spectral indices in the range 144-614 MHz are systematically flatter ($\rm \alpha_{low}\sim$0.55-1.14) than those at higher frequency ($\rm \alpha_{high}\sim$0.75-1.57). However, the spectral distribution within the radio lobes at low frequencies reflects what has been observed at higher frequency (1400-4850 MHz) by \cite{roettiger1994}, that is, an increasing steepening from the inner regions of the lobes towards the lobe edges. This kind of spectral distribution remains very unusual, and has only been observed in one other radio galaxy, Hercules A, which is also claimed to be a restarted source.
\medskip

(ii) By combining the new low-frequency spectral index map with the high-frequency map, we studied the spectral curvature and found values up to 0.7-0.8, especially in the outskirts of the western lobe. This is compatible with old ageing plasma that is not replenished with newly accelerated particles.
\medskip

(iii) We used single-injection models to investigate the age of the source and found that the total source age is equal to $\lesssim$80 Myr. This is consistent with the first-order estimate of the dynamical age of the radio source equal to $<$65 Myr by \cite{kraft2006}.
\medskip

(iv) Considering 3C388 to be a restarted radio galaxy, we estimated the timescales of its duty cycle: $\rm t_{on,1} \gtrsim $30 Myr, $\rm t_{off}\lesssim $ 20 Myr, and $\rm t_{on,2}$>30 Myr for B=$\rm B_{IC}$. These values are consistent with duty cycle estimates derived from other restarted radio galaxies, as well as from multiple generations of X-ray cavities in galaxy clusters.

\medskip

(v) The fitting results using single-injection models (JP and Tribble) over the entire source cannot be rejected with only 68\%\ confidence. The significant number of poorly fitted regions suggests that the spectral shape is not constant throughout the source. This is further highlighted by the colour-colour plot, which shows that the spectra of the western inner lobe better follow a CI model and those of the outer lobe best follow a CIOFF model curve. Mixing of particle populations is the most probable explanation for this behaviour. However, it remains challenging to understand whether this is caused by observational limitations (e.g. insufficient resolution and/or projection effects) or by the intrinsic presence of multiple particle populations.

\medskip

To date, the radio spectral distribution of 3C388 remains a very peculiar case among radio galaxies. However, in the near future it will be much easier to investigate whether more sources with the same characteristics exist. The combination of multi-frequency new-generation instruments and surveys such as the LOFAR Two-metre Sky Survey (LoTSS, \citealp{shimwell2019}), the upgraded Giant Metrewave Radio Telescope (uGMRT, \citealp{gupta2017}), the APERture Tile In Focus (APERTIF, \citealp{oosterloo2009}), the MeerKAT International GHz Tiered Extragalactic Exploration survey (MIGHTEE, \citealp{jarvis2016}) now offer us unprecedented opportunities to perform statistical studies of the spatially resolved spectral properties of restarted radio galaxies, and of radio galaxies in general (\citealp{harwood2016b}, Morganti et al. in prep).

\begin{acknowledgements}
We would like to thank Dharam V. Lal (NCRA–TIFR) for the help provided with the GMRT maps. MB acknowledges support from the ERC-Stg DRANOEL, no 714245 and from INAF under PRIN SKA/CTA ‘FORECaST’. MJH acknowledges support from the UK Science and Technology Facilities Council (ST/R000905/1). MB and IP acknowledge support from the Italian Ministry of Foreign Affairs and International Cooperation (MAECI Grant Number ZA18GR02) and the South African Department of Science and Technology’s National Research Foundation (DSTNRF Grant Number 113121) as part of the ISARP RADIOSKY2020 Joint Research Scheme. LOFAR, the Low Frequency Array designed and constructed by ASTRON (Netherlands Institute for Radio Astronomy), has facilities in several countries, that are owned by various parties (each with their own funding sources), and that are collectively operated by the International LOFAR Telescope (ILT) foundation under a joint scientific policy. We thank the staff of the GMRT that made these observations possible. GMRT is run by the National Centre for Radio Astrophysics of the Tata Institute of Fundamental Research. The National Radio Astronomy Observatory is a facility of the National Science Foundation operated under cooperative agreement by Associated Universities, Inc. This research has made use of the NASA/IPAC Extragalactic Database (NED), which is operated by the Jet Propulsion Laboratory, California Institute of Technology, under contract with the National Aeronautics and Space Administration. This research made use of APLpy, an open-source plotting package for Python hosted at http://aplpy.github.com.
      
\end{acknowledgements}

\bibliographystyle{aa}
\bibliography{3c388_brienza20-saga.bib}

\end{document}